\documentclass[journal]{IEEEtran}
\usepackage{amsfonts}
\usepackage{times}
\usepackage{graphicx}
\usepackage{latexsym}
\usepackage{dsfont}
\usepackage{amssymb}
\usepackage{amsmath}
\usepackage{cite}
\usepackage{verbatim}

\newcommand{\figref}[1]{{Fig.}~\ref{#1}}

\def\bb0{{\mathbb{0}}}

\def\ba{{\mathbf{a}}}
\def\bb{{\mathbf{b}}}

\def\bff{{\mathbf{f}}}
\def\bg{{\mathbf{g}}}
\def\bh{{\mathbf{h}}}

\def\b0{{\mathbf{0}}}

\def\bF{{\mathbf{F}}}
\def\bG{{\mathbf{G}}}
\def\bH{{\mathbf{H}}}

\def\bbC{{\mathbb{C}}}

\def\bbE{{\mathbb{E}}}

\def\cD{\mathcal{D}}
\def\cE{\mathcal{E}}

\def\cH{\mathcal{H}}

\def\cS{\mathcal{S}}

\def\cX{\mathcal{X}}

\def\sf0{{\mathsf{0}}}

\def\nn{\nonumber}

\usepackage{epstopdf}
\usepackage{enumerate}
\usepackage{float}
\usepackage{color}
\usepackage{makeidx}
\usepackage{bm}
\usepackage{cleveref}
\usepackage{url}
\usepackage{subcaption}
\usepackage{balance}
\usepackage{makecell}
\usepackage{booktabs}

\newcommand{\sref}[1]{{Section}~\ref{#1}}

\begin{document}
\bstctlcite{IEEEexample:BSTcontrol}

\title{Digital Twin Aided Massive MIMO CSI Feedback: \\ Exploring the Impact of Twinning Fidelity}

\author{
    \IEEEauthorblockN{Hao Luo, Shuaifeng Jiang, Saeed R. Khosravirad, and Ahmed Alkhateeb}
    \thanks{Part of this work was presented at the IEEE International Conference on Communications (ICC) 2024 \cite{Jiang2024}.}
    \thanks{Hao Luo, Shuaifeng Jiang, and Ahmed Alkhateeb are with the Wireless Intelligence Lab and the School of Electrical, Computer, and Energy Engineering at Arizona State University (email: {h.luo, s.jiang, alkhateeb}@asu.edu). This work was supported in part by the National Science Foundation (NSF) under Grant No. 2048021.}
    \thanks{Saeed R. Khosravirad is with Nokia Bell Laboratories, Murray Hill, NJ 07974, USA (email: saeed.khosravirad@nokia-bell-labs.com).}
}

\maketitle

\begin{abstract}	
    Deep learning (DL) techniques have demonstrated strong performance in compressing and reconstructing channel state information (CSI) while reducing feedback overhead in massive MIMO systems. A key challenge, however, is their reliance on extensive site-specific training data, whose real-world collection incurs significant overhead and limits scalability across deployment sites. To address this, we propose leveraging site-specific digital twins to assist the training of DL-based CSI compression models. The digital twin integrates an electromagnetic (EM) 3D model of the environment, a hardware model, and ray tracing to produce site-specific synthetic CSI data, allowing DL models to be trained without the need for extensive real-world measurements. We further develop a fidelity analysis framework that decomposes digital twin quality into four key aspects: 3D geometry, material properties, ray tracing, and hardware modeling. We explore how these factors influence the reliability of the data and model performance. To enhance the adaptability to real-world environments, we propose a refinement strategy that incorporates a limited amount of real-world data to fine-tune the DL model pre-trained on the digital twin dataset. Evaluation results show that models trained on site-specific digital twins outperform those trained on generic datasets, with the proposed refinement method effectively enhancing performance using limited real-world data. The simulations also highlight the importance of digital twin fidelity, especially in 3D geometry, ray tracing, and hardware modeling, for improving CSI reconstruction quality. This analysis framework offers valuable insights into the critical fidelity aspects, and facilitates more efficient digital twin development and deployment strategies for various wireless communication tasks.
\end{abstract}

\section{Introduction}
    Multiple-Input and Multiple-Output (MIMO) communications have been widely recognized as an important feature in the current and next-generation wireless systems. By employing large antenna arrays, MIMO systems can offer promising spatial multiplexing and beamforming gains~\cite{Foschini1998, Telatar1999, Love2005, Marzetta2010, Ngo2013}. To design the precoding vectors and fully exploit these gains, the base station (BS) requires accurate channel state information (CSI) for the downlink channel. Typically, the downlink CSI acquisition process consists of three steps~\cite{Love2008}. (i) The BS transmits downlink pilot signals to the user equipment (UE). (ii) The UE estimates the CSI of the downlink channel based on these pilot signals. (iii) The UE sends the estimated CSI back to the BS via the uplink control channel. However, with traditional CSI feedback methods~\cite{Visotsky2001, Narula1998, Lau2004, Roh2006}, the associated downlink pilot and uplink feedback overhead increases rapidly with the number of antennas, posing challenges for MIMO systems to further scale their gains with more antennas.

    Deep learning (DL) has recently found widespread application across various MIMO communication tasks, including CSI feedback~\cite{Wen2018, Jiang2023b,Li2019}, codebook learning~\cite{Zhang2022}, robust beamforming design~\cite{Zhao2025}, and many others. Among these, DL-aided CSI feedback is a promising solution for mitigating the CSI acquisition overhead challenge in MIMO systems. It is worth noting that, depending on the frequency band and array architecture, CSI feedback can be represented in different forms and acquired through various methods. For instance, in massive MIMO systems, prior work has leveraged DL models to compress the estimated channel at the UE, achieving higher efficiency compared to classical CSI compression approaches~\cite{Wen2018}. In millimeter-wave (mmWave) MIMO systems, DL models have been employed to predict current or future beams based on prior information, e.g., multi-modal sensing, thereby reducing the downlink beam training overhead~\cite{Jiang2023b}. Furthermore, for MIMO systems with hybrid digital-analog architectures, DL models have been used to learn compressive sensing measurement vectors, which helps to reduce the downlink pilot signal overhead~\cite{Li2019}. However, these DL-aided wireless communication approaches generally require extensive real-world CSI data for training. In deep learning, the training data distribution needs to align well with the test data distribution for the DL model to achieve high performance. Changes in the wireless environment, e.g., new building deployments or foliage growth, can significantly alter the CSI distribution. This necessitates extensive collection of new real-world datasets. From a long-term perspective, this considerable overhead impedes the scalability of DL-aided wireless communications across numerous sites.
    
    To address this challenge, in this paper, we investigate the use of site-specific digital twins~\cite{Alkhateeb2023} to mitigate the real-world data collection overhead of DL-aided CSI feedback for massive MIMO systems. Specifically, we adopt a physics-based digital twin construction approach\footnote{Another digital twin construction approach is data-driven digital twin construction, which uses generative models to predict site-specific channel~\cite{Lee2024} or system performance~\cite{He2023}. However, similar to the DL-aided wireless communications, this approach also faces the challenge of extensive real-world data collection overhead, especially when environmental changes occur.}. This involves building a digital replica of the communication environment and hardware characteristics, and then simulating signal propagation using ray tracing software. We can use digital twins to generate offline, site-specific CSI data for training DL-based CSI compression models, which closely resembles real-world data. Notably, the cost of a physics-based digital twin is mainly a one-time investment. While initial real-world measurements are needed to construct the foundational digital replica, this upfront cost can be significantly amortized across various communication tasks and even across different industries, e.g., smart cities and intelligent transportation systems~\cite{Jafari2023}. Also, physics-based digital twin construction offers robust adaptability to changes in wireless environments. Essentially, the digital twin can be efficiently updated by partially revising the 3D model and/or the EM properties of certain objects to reflect environmental changes, thus avoiding the need for a complete reconstruction. Furthermore, we also study the impact of digital twin fidelity on communication performance. We propose decomposing digital twin fidelity into four aspects: 3D geometry, electromagnetic (EM) material, ray tracing, and hardware modeling. We consider DL-aided massive MIMO CSI feedback as a case study to explore how these aspects of digital twin fidelity affect the performance of DL models. The same analysis framework can be applied to other digital twin aided wireless communication tasks.

    \subsection{Related Work}
        \textbf{Digital twins aided MIMO communications.} MIMO communication systems can leverage large antenna arrays to enhance beamforming and spatial multiplexing gains. However, increasing the number of antennas largely raises the overhead associated with CSI estimation and feedback. To address this challenge, deep learning has been explored as a potential solution for various scenarios, such as massive MIMO CSI feedback~\cite{Wen2018}, mmWave beam prediction~\cite{Jiang2023b}, and compressive sensing for hybrid precoding~\cite{Li2019}. Despite its promise, the need for extensive real-world CSI data remains a major challenge in training DL models, creating a ``chicken-and-egg'' problem. Recently, site-specific digital twins~\cite{Alkhateeb2023} have been envisioned as a promising source of synthetic data to facilitate DL model training for MIMO communications. Site-specific digital twins utilize high-precision 3D maps and ray-tracing simulators to approximate the real-world communication environment, which has shown great potential in the literature. For instance, the authors in~\cite{Jiang2023a} proposed using digital twins to assist the DL-based beam prediction problem, where the DL model is trained on synthetic data generated by the digital twin. Also, in \cite{Luo2024}, a digital twin was used to aid compressive sensing in hybrid MIMO systems, where the measurement vectors learned from the site-specific synthetic data can effectively capture the promising directions of real-world channels. For massive MIMO systems, in \cite{Jiang2024}, we proposed digital twin aided CSI feedback, where a digital twin is used to generate synthetic CSI data for training the DL-based CSI compression model. Following our initial work, the authors in~\cite{Huang2025} explored the effectiveness of digital twins for CSI compression by evaluating them with real-world CSI samples. However, a comprehensive analysis of digital twin fidelity and a robust solution for real-world data mismatch have not been addressed. In this paper, we develop an analysis framework to study the fidelity of the digital twin and its impact on the performance of DL-aided wireless communication tasks, with DL-aided CSI compression as a case study. Moreover, we propose a novel data selection approach for online model refinement, which can effectively enhance the model's performance with a small amount of real-world data.

        \textbf{Deep learning aided massive MIMO CSI feedback.} Deep learning has been extensively studied in recent years to enhance CSI compression and feedback in massive MIMO systems~\cite{Wen2018,Wang2018,Tolba2020,Guo2020,Mismar2024,Mashhadi2020,Ji2021,Lu2021,Singh2024,Cui2024,Liu2021,Chen2021,Cui2022a,Cui2022b}. The key idea is to employ an autoencoder neural network, consisting of an encoder and a decoder, for CSI compression and reconstruction. Specifically, the UE first employs the encoder to compress the full CSI into a lower-dimensional form before the feedback. Then, the BS uses the decoder to reconstruct the CSI from its compressed representation. The authors in \cite{Wen2018} were the first to propose using DL to solve the CSI compression and recovery problem. Since then, various DL model designs have been introduced to improve the performance. Specifically, in \cite{Wang2018}, the authors aimed to utilize the temporal correlation in the wireless channels and proposed a recurrent neural network (RNN) based solution for time-varying channels. In \cite{Tolba2020}, the authors used generative adversarial networks (GANs) to improve CSI recovery performance. Additionally, practical considerations have been taken into account to facilitate the deployment of DL-based CSI feedback in real-world systems. For instance, in \cite{Guo2020}, the authors introduced a multiple-rate neural network to compress and quantize the CSI with varying numbers of feedback bits. In \cite{Mismar2024}, the authors studied DL-based CSI compression as an element of a neural receiver with an adaptive compression ratio, and they discussed several practical considerations. In \cite{Lu2021}, a lightweight NN architecture was proposed for CSI feedback, in which the model weights are quantized with one bit. Similarly, \cite{Singh2024} and \cite{Cui2024} respectively proposed model pruning and knowledge distillation approaches to reduce the model size and computational complexity. In \cite{Liu2021} and \cite{Chen2021}, the authors explored the implicit CSI feedback, in which the precoder matrix is fed back instead of the full CSI. In \cite{Cui2022b}, the authors introduced an unsupervised online learning approach to adapt to changing channel statistics. However, all these approaches require real-world CSI data for training the DL models, leading to significant overhead in data collection. In this paper, we explore the use of site-specific digital twins to generate CSI data that closely resembles the real-world data, which helps reduce the data collection overhead.

    \subsection{Contributions}
        In this work, we propose a digital twin aided CSI compression method, where the digital twin generates site-specific synthetic CSI data that is used to train a deep learning model. The proposed approach achieves high CSI compression performance and significantly reduces the costly real-world data collection overhead. To the best of our knowledge, this is the first work to employ site-specific digital twins for CSI compression. The main contributions of this paper are summarized as follows.
        \begin{itemize}
            \item We propose a novel approach that utilizes site-specific digital twins to facilitate the DL model training for massive MIMO CSI feedback. The digital twin generates site-specific synthetic CSI data based on the EM 3D model, hardware model, and ray tracing simulation, which can then be used to train the DL model without the need for real-world data collection.
            \item We develop a novel framework for quantifying and analyzing the fidelity of site-specific digital twins, decomposing it into four key aspects: 3D geometry, EM material, ray tracing, and hardware model. This framework enables a systematic evaluation of how each fidelity aspect impacts the performance of DL-aided wireless communication tasks. While our case study focuses on DL-aided CSI feedback, this analysis framework is applicable to other digital twin aided wireless communications.
            \item We propose an online data selection and model refinement approach to fine-tune the DL model trained on the digital twin data. The proposed approach can effectively mitigate the mismatch between the real-world and digital twin CSI distributions, which enhances the DL performance using a limited amount of real-world data. As a result, our approach can achieve high CSI compression performance with significantly reduced real-world data collection overhead.
            \item To evaluate the proposed approach, we construct a digital twin dataset based on a ray tracing scenario that captures the geometry of downtown Boston, which serves as the real-world environment. Based on the original scenario, we build digital twin scenarios with varying fidelity levels by following a structured pipeline, which enables the study of the impact of digital twin fidelity.
        \end{itemize}
        In the simulation, we conduct extensive experiments to evaluate the performance of the proposed approach using the digital twin dataset. First, the model trained on the site-specific digital twin data achieves high performance when tested in the real-world deployment. Notably, it outperforms the model trained on the generic dataset in terms of both CSI recovery and feedback overhead. Second, the proposed online data selection and model refinement approach can effectively enhance the model pre-trained on the digital twin data with a small amount of real-world data. Compared to the model trained solely on the same amount of real-world data, the proposed approach achieves a significant performance gain, demonstrating the effectiveness of the digital twin in reducing the data collection overhead. Third, through a study on the impact of digital twin fidelity, we identify that 3D geometry, ray tracing, and hardware model are the key factors that greatly influence the performance of CSI reconstruction. Lastly, we discuss the comparison between real-world channel estimation noise and digital twin imperfections, which showcases the potential of the digital twin in practical scenarios.

    The remainder of the paper is organized as follows. \sref{sec:System and Channel Models} introduces the system and channel models. \sref{sec:Deep Learning Based CSI Feedback} presents the DL-based CSI feedback and the objective of the DL model training. \sref{sec:Digital Twin for Wireless Communications} elaborates the concept of site-specific digital twin and defines the fidelity of the digital twins. \sref{sec:Digital Twin Aided CSI Feedback} presents the proposed digital twin aided CSI feedback approach. \sref{sec:Simulation Setup} includes the simulation setup and dataset generation. The evaluation results are discussed in \sref{sec:Evaluation Results}. Finally, \sref{sec:Conclusion} concludes the paper, and \sref{sec:Future_Directions} discusses future research directions.

\begin{figure*}[t]
	\centering
	\includegraphics[width=1\linewidth]{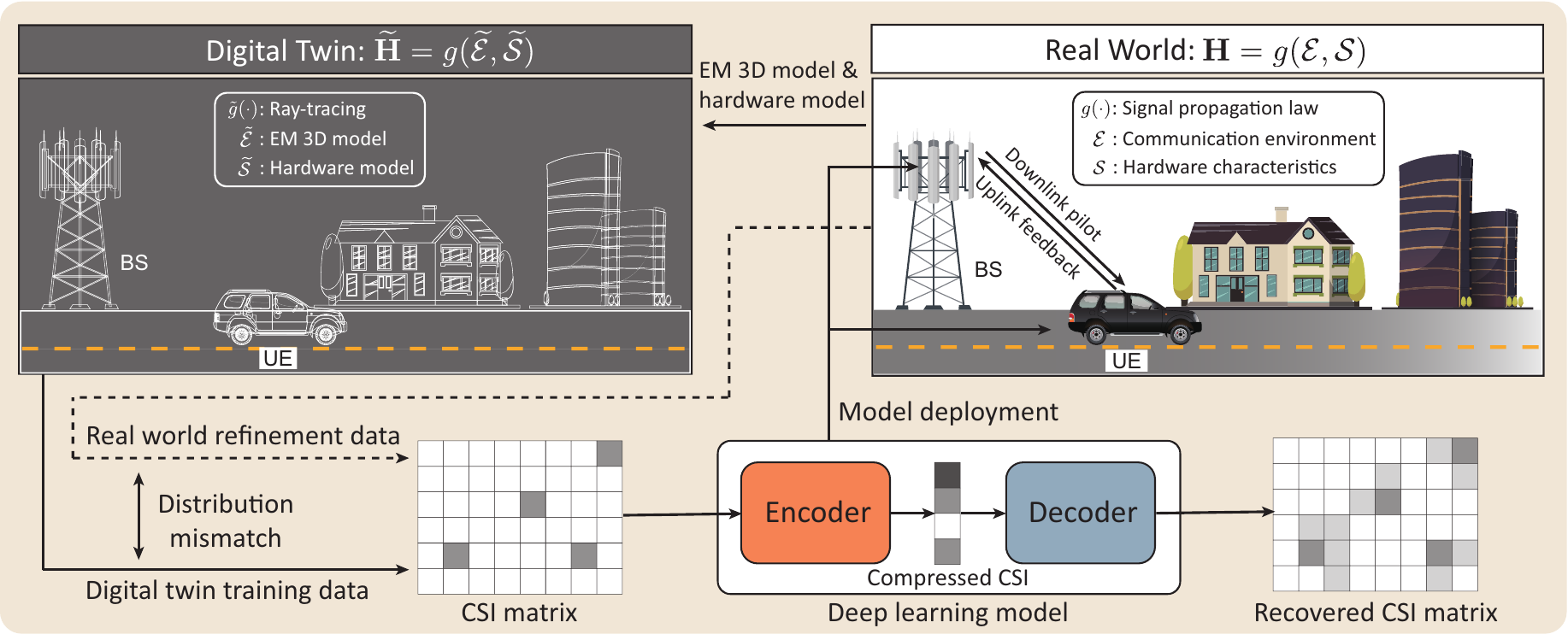}
	\caption{This figure shows the key idea of utilizing a digital twin to train the DL CSI model. A small amount of real-world data can then be used to compensate for the mismatch between the digital twin and real-world CSI data, and refine the DL model to achieve higher performance.}
	\label{fig:key_idea}
\end{figure*}

\section{System and Channel Models} \label{sec:System and Channel Models}
    We adopt a frequency division duplex (FDD) single-cell system in which a BS equipped with $N_t$ antennas communicates with $U$ single-antenna UEs. The system employs OFDM modulation across $K$ subcarriers. Let $x_{k,u} \in \bbC$ represent the downlink complex symbol for the $u^\text{th}$ UE on the $k^\text{th}$ $(k\in\{1,\hdots,K\})$ subcarrier with unit average power, i.e.,  $\bbE[x_{k,u}^\textrm{H}x_{k,u}]=1$. The received signal for the $u^\text{th}$ UE on the $k^\text{th}$ subcarrier can then be expressed as
    \begin{align}
        y_{k,u} = \bh_{k,u}^\textrm{H}\bff_{k,u} x_{k,u} + \sum_{v \neq u} \bh_{k,u}^\textrm{H}\bff_{k,v} x_{k,v} + n_{k,u},
    \end{align}
    where $\bh_{k,u} \in \bbC^{N_t\times 1}$ denotes the downlink channel vector, and $\bff_{k,u} \in \bbC^{N_t\times 1}$ denotes the precoding vector, which satisfies the sum-power constraint $ \sum_{u=1}^U \|\bff_{k,u}\|_2^2 \leq P$. $n_{k,u} \in \bbC$ is the additive Gaussian noise with zero mean and variance $\sigma^2_n$. We employ a block-fading wideband geometric channel model, where the delay-d channel vector $\bh_{d,u}$ can be expressed as
    \begin{align}\label{eq:delayd_channel}
        \bh_{d,u} = \sum_{l=1}^{L}\alpha_{l,u} p(dT_S-\tau_{l,u})\ba(\phi_{l,u}, \theta_{l,u}),
    \end{align}
    where $p(\tau)$ is the pulse shaping function, representing a $T_S$-spaced signaling evaluated at $\tau$ seconds. The channel comprises $L$ paths, with $\alpha_{l,u}$, $\tau_{l,u}$, $\phi_{l,u}$, and $\theta_{l,u}$ indicating the complex gain, propagation delay, azimuth and elevation angles of departure (AoD) of the $l^\text{th}$ path, respectively. The term $\ba(\phi_{l,u}, \theta_{l,u})$ represents the BS array response, which varies based on the array geometry. The frequency domain channel vector on the $k^\text{th}$ subcarrier can then be written as
    \begin{align}\label{eq:frequency_channel}
        \bh_{k,u} = \sum_{d=0}^{D-1}\bh_{d,u} \exp\left(-j\frac{2\pi k}{K}d\right), 
    \end{align}
    where $D$ represents the maximum channel delay. The sum rate for the downlink data transmission can then be expressed as
    \begin{equation} \label{eq:sum_rate}
        R = \frac{1}{K} \sum_{k=1}^K \sum_{u=1}^U  \log_2\left(1+\frac{\left|\bh_{k,u}^\textrm{H}\,\bff_{k,u}\right|^2}{\sum_{v \neq u} \left|\bh_{k,u}^\textrm{H}\,\bff_{k,v}\right|^2 + \sigma_n^2}  \right).
    \end{equation}
    Let $\bH_u = [\bh_{1,u}, \hdots, \bh_{K,u}] \in \bbC^{N_t \times K}$ denote the downlink channel matrix of the $u^\text{th}$ UE obtained by combining the $K$ channel vectors. To achieve high sum rates $R$, the BS relies on the precise CSI of the downlink channel to design the precoding vectors $\bff_{k,u}$. In FDD systems, this CSI is acquired through downlink training and uplink feedback processes. These processes begin with the BS sending predefined downlink pilot signals to the UE. Using these pilots, the UE estimates the CSI and subsequently transmits the compressed CSI back to the BS over the uplink feedback channel. The reconstructed CSI matrix at the BS is denoted by $\widehat{\bH}_u = [\widehat{\bh}_{1,u}, \dots, \widehat{\bh}_{K,u}] \in \bbC^{N_t \times K}$. In this work, we assume the uplink feedback channel is noiseless, and that imperfections in the CSI feedback arise from errors during the estimation and compression processes at the UE.

\section{Deep Learning Based CSI Feedback} \label{sec:Deep Learning Based CSI Feedback}
    In this section, we introduce the DL-based CSI feedback approach for massive MIMO systems. To enable spatial multiplexing and beamforming gains, the UE needs to feed back the estimated CSI matrix to the BS for precoder design. However, since the size of the CSI matrix increases in proportion to the number of antennas and subcarriers, transmitting it directly results in significant wireless resource overhead. This feedback overhead can substantially lower system performance due to the limited channel coherence time. To tackle this, the CSI is usually compressed at the UE prior to transmission. Since each UE performs the compression individually, we will drop the subscript $u$ from the CSI matrix for ease of notation. Let $f_\textrm{enc}(\cdot)$ represent the CSI compression function, or encoder. The compressed CSI, denoted as $\bg \in \bbC^{M \times 1}$, is written as 
    \begin{equation}
        \bg = f_\textrm{enc}(\bH),
    \end{equation}
    where the size of the compressed CSI is less than that of the original CSI matrix, i.e., $M< N_tK$. The compressed CSI $\bg$ is then transmitted to the BS with less feedback overhead. At the BS, the CSI matrix $\widehat{\bH}$ is reconstructed from the compressed CSI $\bg$, which can be written as
    \begin{align}
        \widehat{\bH} &= f_\textrm{dec}(\bg) = f_\textrm{dec}\big(f_\textrm{enc}(\bH)\big),
    \end{align}
    where $f_\textrm{dec}(\cdot)$ denotes the CSI recovery function, or decoder.

    In this paper, we aim to develop CSI compression and reconstruction functions using DL models, ensuring that the CSI recovered at the BS aligns closely with the CSI estimated at the UE for a given channel distribution \mbox{$\bH \sim \cH$}. To assess the discrepancy between the estimated CSI and the reconstructed CSI, we use the normalized mean square error (NMSE), which is defined as follows:
    \begin{equation}\label{eq:nmse}
        \mathsf{NMSE}(\bH, \widehat{\bH}) = \frac{\|\bH - \widehat{\bH}\|_F^2}{\|\bH\|_F^2},
    \end{equation}
    where $\|\cdot\|_F$ denotes the Frobenius norm.
    Let $f_\textrm{enc}(;\Theta_\textrm{enc})$ and $f_\textrm{dec}(;\Theta_\textrm{dec})$ denote the DL models for the CSI compression and recovery functions, where $\Theta_\textrm{enc}$ and $\Theta_\textrm{dec}$ are the trainable parameters. Our goal is to develop DL models that minimize the channel reconstruction NMSE. The optimal DL models can be expressed as
    \begin{align}\label{eq:optimal_ml_model}
        & f^\star_\textrm{enc}(;\Theta_\textrm{enc}^\star), \, f^\star_\textrm{dec}(;\Theta_\textrm{dec}^\star) \nn\\
        =\,& \underset{\substack{f_\textrm{enc}(;\Theta_\textrm{enc})\\ f_\textrm{dec}(;\Theta_\textrm{dec})}}{\arg\min} \ \bbE_{\bH \sim \cH}\bigg\{  \mathsf{NMSE}\Big[\bH, f_\textrm{dec}\big(f_\textrm{enc}(\bH)\big)\Big]\bigg\}.
    \end{align}

    In practice, the true CSI distribution is approximated using a dataset of CSI matrices. Let \mbox{$\cD = \{\bH_1, \hdots, \bH_{N_d}\}$} denote the CSI dataset, which includes $N_d$ data points sampled from $\cH$. The DL models can then be trained end-to-end by minimizing the following loss function:
    \begin{align}\label{eq:ml_loss}
        L(\Theta_\textrm{enc}, \Theta_\textrm{dec}, \cD) = \frac{1}{N_d}\sum_{n=1}^{N_d} \mathsf{NMSE}\Big[\bH_n, f_\textrm{dec}\big(f_\textrm{enc}(\bH_n)\big)\Big].
    \end{align}
    A key challenge with the DL-based CSI feedback is the need for real-world training data. To ensure robust performance, the DL model typically requires a dataset that captures the full diversity of the CSI distribution. It is worth noting that the CSI distribution is strongly tied to the communication environment and differs across various sites. This variation leads to a substantial overhead in collecting real-world CSI data, posing a major barrier to scaling DL-based CSI feedback across numerous communication sites.

\section{Site-Specific Digital Twins} \label{sec:Digital Twin for Wireless Communications}
    To reduce the overhead of real-world data collection, site-specific digital twins~\cite{Alkhateeb2023} offer a promising solution by generating synthetic CSI data that closely resembles its real-world counterpart. In this section, we introduce the concept of site-specific digital twins and outline an exemplary process for constructing a digital twin. Moreover, we define the fidelity of digital twins by breaking it down and examining its various aspects. An illustration of the key idea is shown in \figref{fig:key_idea}.

    \subsection{Key Components}
        The real-world communication channel is primarily determined by three factors: 
        \begin{enumerate}[(i)]
            \item \textbf{Communication environment} $\cE$ includes the positions, orientations, dynamics, and shapes of the BS, UE, and surrounding objects (e.g., reflectors and scatterers).
            \item \textbf{Signal propagation law} $g(\cdot)$ governs how signals propagate through the environment.
            \item \textbf{Hardware characteristics} $\cS$ defines the physical properties and impairments of communication hardware.
        \end{enumerate}
        Based on these three components, the communication channel can be expressed as
        \begin{align}
            \bH = g(\cE, \cS).
        \end{align}
        However, obtaining an accurate representation of the communication environment $\cE$ and precisely defining the signal propagation law $g(\cdot)$ are challenging in complex environments. To address this, we can develop a site-specific digital twin that provides an approximation of both $\cE$ and $g(\cdot)$. In particular, the digital twin utilizes an EM 3D model to depict the communication environment and applies ray tracing to simulate the signal propagation law. Additionally, it integrates hardware properties and impairments $\mathcal{S}$ by combining empirical measurements with modeling techniques. Below, we provide more details about the core components of a digital twin.

        \textbf{EM 3D model.} The EM 3D model $\widetilde{\cE}$ provides information about the positions, orientations, dynamics, and shapes of the BS, UE, and surrounding objects (e.g., reflectors and scatterers). Additionally, it includes material properties of the objects, such as permittivity and conductivity, which influence the interaction between EM waves and objects in the environment. This information can be obtained from sources such as architectural designs, remote sensing, and field measurements.

        \textbf{Ray tracing.} The ray tracing $\widetilde{g}(\cdot)$ simulates signal propagation paths between each transmit-receive antenna pair based on the geometry and material properties from the EM 3D model. The simulation accounts for various propagation effects, including reflection, diffraction, and diffuse scattering. As a result, it generates key path parameters such as complex path gain, propagation delay, and angles of arrival and departure. These parameters ensure spatially consistent channel synthesis that accurately aligns with the environment’s geometry.

        \textbf{Hardware model.} The hardware model $\widetilde{\cS}$ captures various essential characteristics of the communication system, including antenna type, radiation pattern, and array geometry. Additionally, it includes transceiver impairments such as power-amplifier distortion, phase noise, and quantization noise, which can impact signal quality and system performance. This model can be developed using manufacturer-provided hardware specifications, theoretical models, and empirical measurements.
        
        Given the components mentioned above, the site-specific digital twin generates synthetic CSI, defined as:
        \begin{align}
        \widetilde{\mathbf{H}} = \widetilde{g}(\widetilde{\mathcal{E}}, \widetilde{\mathcal{S}}),
        \end{align}
        which approximates real-world CSI. If the synthetic CSI closely aligns with real-world CSI, it can be utilized to train the DL model for CSI reconstruction. This approach can reduce or even eliminate the need for extensive real-world CSI data collection.

        \subsection{Digital Twin Construction} \label{sec:Digital Twin Construction}
        With the concept and key components of site-specific digital twins defined, we now focus on their practical implementation. Various methods exist for constructing a digital twin, and in this section, we introduce a specific pipeline designed for generating synthetic, site-specific CSI data. 
        
        \textbf{Point cloud data collection.} A widely adopted approach for constructing a 3D geometry model is to use point cloud data, which is often collected using LiDAR. LiDAR is a remote sensing technology that measures distances to objects by emitting laser light and analyzing the reflected signals with a sensor. LiDAR data enables the creation of a detailed 3D point cloud representing the communication environment.

        \textbf{Mesh reconstruction from 3D point cloud.} The 3D point cloud serves as the basis for generating the 3D geometry model. During this process, a mesh reconstruction algorithm~\cite{Kazhdan2013,Shen2021} is employed to convert the point cloud into a 3D mesh. The mesh consists of vertices, edges, and faces, which define the shapes and structures of the objects. However, the mesh reconstruction process is generally imperfect, and distortions often arise from algorithmic limitations or insufficient point cloud data.

        \textbf{EM material and hardware modeling.} Once the 3D mesh is generated, the next step is to assign material properties to the objects. This involves a semantic segmentation step on either the 3D point cloud~\cite{Zhang2019} or the 3D mesh~\cite{Kalogerakis2010} to identify different parts of the objects and assign material properties such as permittivity and conductivity. These properties can be obtained from material specifications or measurements. Additionally, the hardware model is developed to capture communication system characteristics, including antenna types, radiation patterns, and transceiver impairments.

        \textbf{Ray tracing for CSI data generation.} The final step in the digital twin construction is to generate synthetic CSI data using ray tracing. Given the EM 3D geometry model and hardware model, a ray tracer~\cite{Remcom,Hoydis2023} is used to simulate signal propagation paths between the transmitter and receiver. The ray tracer generates channel path parameters, including path gain, propagation delay, and angles of arrival and departure. These parameters can be used to generate channels that are spatially consistent with the environment's geometry.
        
    \subsection{Fidelity of Digital Twins}
        In practical scenarios, the digital twin may not perfectly replicate the real-world communication environment. Such imperfections can result in a mismatch between the synthetic and real-world CSI data distributions. Thus, quantifying the fidelity of the digital twin is crucial for understanding its impact on DL model performance. In the existing literature, however, how to quantify the fidelity of the digital twin is still an open question. In this paper, we develop a method to quantify digital twin fidelity by decomposing it into four aspects: 3D geometry, EM material, ray tracing, hardware model. Overall, the fidelity of a digital twin is multi-dimensional, with each aspect measured by distinct metrics. Below, we elaborate on these aspects of digital twin fidelity in more detail.

\begin{figure}[t]
            \centering
            \includegraphics[width=1\linewidth]{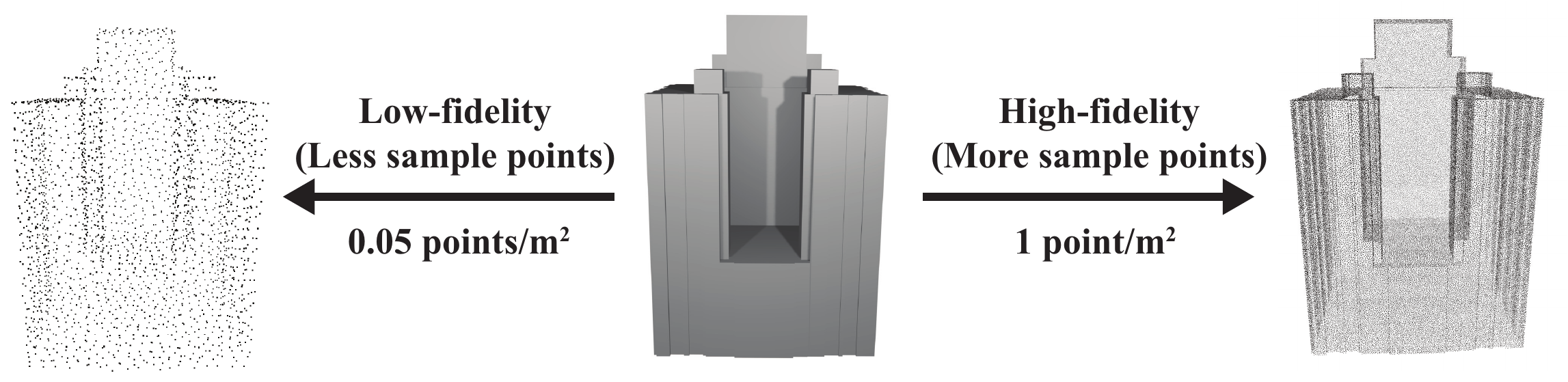}
            \caption{This figure illustrates the point cloud generated at different sampling densities, where sampling density refers to the number of points per square meter. A higher sampling density captures more details of the building model, resulting in greater geometric fidelity after mesh reconstruction.}
            \label{fig:sample_density_vs_fidelity}
        \end{figure}

        \textbf{3D geometry.} The fidelity of the 3D geometry model relates to how accurately it represents real-world geometry. If we consider the real-world geometry as a reference, the fidelity of the digital twin can be quantified by measuring the similarity between these two models. Typically, there is no direct one-to-one correspondence between the vertices of the two models. Instead, researchers commonly sample points from both models and compare these sampled points to evaluate similarity. For instance, the F1-score~\cite{Hanocka2020} is a widely used metric for evaluating the similarity between two point clouds. Given the sampling points of the original building model $\cX$ and the reconstructed building model $\widehat{\cX}$, the F1-score is defined as
        \begin{equation} \label{eq:f1_score}
            F(\tau) = \frac{2P(\tau)R(\tau)}{P(\tau)+R(\tau)},
        \end{equation}
        where $P(\tau)$ and $R(\tau)$ are the precision and recall at threshold $\tau$, respectively. The precision and recall are defined as
        \begin{align}
            P(\tau) &= 100 \times \frac{\sum_{x \in \cX} \left[ d_{x \rightarrow \widehat{x}}<\tau \right]}{|\cX|},\\
            R(\tau) &= 100 \times \frac{\sum_{\widehat{x} \in \widehat{\cX}} \left[ d_{\widehat{x} \rightarrow x}<\tau \right]}{|\widehat{\cX}|},
        \end{align}
        where $d_{x \rightarrow \widehat{x}}$ and $d_{\widehat{x} \rightarrow x}$ are the distances from the points in $\cX$ to the nearest points in $\widehat{\cX}$ and vice versa, respectively. The F1-score is a metric that computes the harmonic mean between the precision and recall to measure the similarity between the original and reconstructed building models. We choose the threshold $\tau$ by following the method in \cite{Hanocka2020}\footnote{The original building model is sampled twice, and the one-directional distance is computed between samples, similar to the calculation in the F1-score. The threshold $\tau$ is set to the $100^\text{th}$ percentile of the computed distances, i.e., the maximum distance between the two ground-truth point clouds.}. In practice, however, we may not have access to building models that perfectly match real-world geometry. Instead, we can use the data collection overhead of the point cloud as a proxy for the fidelity of the 3D geometry model. This overhead can be quantified by the number of points per square meter. As illustrated in \figref{fig:sample_density_vs_fidelity}, a higher sampling density captures more details of the building model, resulting in greater fidelity. Furthermore, our evaluation will demonstrate that, for a given mesh reconstruction algorithm, the number of sampling points is positively correlated with both the F1-score and the performance of the DL model.

        \textbf{EM material.} Once material properties are assigned to each object in the 3D geometry model, the fidelity of the EM material can be evaluated by comparing the material properties of the real-world and digital twin objects. Material properties are typically characterized by the relative permeability $\mu_r$ and the complex relative permittivity $\varepsilon_c$~\cite{Hoydis2024}. We assume the materials are non-magnetic, where the relative permeability is equal to the vacuum permeability, i.e., $\mu_r = 1$. Then, the material property can be represented by the complex relative permittivity $\varepsilon_c$ as
        \begin{equation}
            \varepsilon_c = \varepsilon_r - j\frac{\sigma}{\varepsilon_0\omega},
        \end{equation}
        where $\varepsilon_0$ is the vacuum permittivity, and $\omega=2\pi f$ is the angular frequency. The complex relative permittivity, $\varepsilon_c$, depends on the real-valued relative permittivity, $\varepsilon_r \geq 1$, and the conductivity, $\sigma \geq 0$. The EM material fidelity is then quantified by using the relative difference between the real-world and digital twin material properties, expressed through these two parameters. This relative difference is defined as
        \begin{equation}
            \Delta \varepsilon_r = \frac{|\bar{\varepsilon}_r - \widetilde{\varepsilon}_r|}{\bar{\varepsilon}_r}, \quad \Delta \sigma = \frac{|\bar{\sigma} - \widetilde{\sigma}|}{\bar{\sigma}},
        \end{equation}
        where $\bar{\varepsilon}_r$ and $\bar{\sigma}$ are real-world material properties, and $\widetilde{\varepsilon}_r$ and $\widetilde{\sigma}$ are digital twin material properties. Thus, EM material fidelity is a two-dimensional measure that influences the modeling of EM wave interactions with objects in ray tracing.

        \textbf{Ray tracing.} To define the fidelity of ray tracing, we assume that the ray tracing algorithm remains identical throughout the evaluation. This accounts for the fact that different ray tracers may rely on varying assumptions. For instance, the current version of Sionna~\cite{Hoydis2023} assumes that if diffraction is present in a path, it is the only interaction, whereas Wireless Insite~\cite{Remcom} allows multiple interactions in a single path. Such differences in assumptions can lead to variations in the generated channel, making it challenging to define ray tracing fidelity. Grounded in this assumption, the fidelity of ray tracing can be quantified by the configurations of the ray tracing algorithm, including the number of reflections, diffractions, and diffuse scattering. These configurations directly affect the existence of a path between the transmitter and receiver, thereby influencing the channel synthesis.

        \textbf{Hardware model.} Modeling hardware characteristics is crucial for accurately simulating communication systems. These characteristics can arise from various sources, such as antenna element mismatches, transceiver phase noise, power amplifier non-linearities, and DAC/ADC non-idealities. Given that hardware modeling is a broad and complex topic, this work focuses on one critical aspect: the field of view (FoV) of the BS antenna array. We defer a more comprehensive study on hardware modeling to future work. The FoV is a critical factor for channel synthesis, since any mismatch can directly impact whether a path is observable, leading to a mismatch between synthetic and real-world CSI distributions.

\section{Digital Twin Aided CSI Feedback} \label{sec:Digital Twin Aided CSI Feedback}
    In this section, we propose a digital twin aided CSI feedback approach to reduce the real-world data collection overhead for the DL-based CSI feedback. The proposed approach consists of two phases: (i) direct generalization and (ii) model refinement. The direct generalization phase involves training the DL model on digital twin synthetic data. The model refinement phase further refines the DL model using a small amount of real-world data. An illustration of the proposed approach is shown in \figref{fig:key_idea}.

    \subsection{Problem Formulation}
        Let $\cH$ represent the CSI distribution in the real world and $\widetilde{\cH}$ represent the CSI distribution in the digital twin. Our goal is to use the digital twin CSI distribution to develop a DL model that performs comparably to a DL model obtained using the real-world CSI distribution. Let $\tilde{f}_\textrm{enc}(;\widetilde{\Theta}_\textrm{enc})$ and $\tilde{f}_\textrm{dec}(;\widetilde{\Theta}_\textrm{dec})$ denote the DL models trained on the digital twin CSI distribution $\widetilde{H}$. The objective can be mathematically formulated by
        \begin{align}\label{eq:objective}
            \underset{\substack{\tilde{f}_\textrm{enc}(;\widetilde{\Theta}_\textrm{enc})\\ \tilde{f}_\textrm{dec}(;\widetilde{\Theta}_\textrm{dec})}}{\min} \left| L(\widetilde{\Theta}_\textrm{enc}, \widetilde{\Theta}_\textrm{dec}, \cH)-L(\Theta_\textrm{enc}^\star, \Theta_\textrm{dec}^\star, \cH)\right|.
        \end{align}
        The goal is to minimize the discrepancy between the loss of the DL model trained on digital twin data and that of the model trained on real-world data. 

    \subsection{Direct Generalization} \label{sec:Direct Generalization}
        From a machine learning perspective, training a DL model on digital twin synthetic data to achieve high performance on real-world data is a domain adaptation problem \cite{Zhao2020}. Here, the digital twin synthetic data distribution is the source domain, and the real-world data distribution is the target domain. In \cite{Mansour2009}, the authors showed that, for any mapping function, i.e., the CSI compression and recovery models $\widetilde{f}_\textrm{enc}(;\widetilde{\Theta}_\textrm{enc})$ and $\widetilde{f}_\textrm{dec}(;\widetilde{\Theta}_\textrm{dec})$ in our context, the following bound holds:
        \begin{align}\label{eq:bound}
            &\left| L(\widetilde{\Theta}_\textrm{enc}, \widetilde{\Theta}_\textrm{dec}, \cH)-L(\Theta_\textrm{enc}^\star, \Theta_\textrm{dec}^\star, \cH)\right| \nonumber\\
            \leq \ &L(\widetilde{\Theta}_\textrm{enc}, \widetilde{\Theta}_\textrm{dec}, \widetilde{\cH}) + \mathrm{disc}(\cH, \widetilde{\cH}) + \epsilon.
        \end{align}
        Here, $\mathrm{disc}(\cH, \widetilde{\cH}) > 0 $ represents the discrepancy between the two distributions $\cH$ and $\widetilde{\cH}$ while $\epsilon > 0$ is a constant determined by $\cH$, $\widetilde{\cH}$, and the function classes of $f_\textrm{enc}$ and $f_\textrm{dec}$. As both $\mathrm{disc}(\cH, \widetilde{\cH})$ and $\epsilon$ remain constant for given $\cH$ and $\widetilde{\cH}$, the upper bound of \eqref{eq:objective} can be reduced by minimizing $L(\widetilde{\Theta}_\textrm{enc}, \widetilde{\Theta}_\textrm{dec}, \widetilde{\cH})$, which corresponds to the DL model's loss on the digital twin synthetic data. Therefore, the DL model can be obtained by minimizing the following loss function:
        \begin{align}\label{eq:ml_loss2}
            L(\widetilde{\Theta}_\textrm{enc}, \widetilde{\Theta}_\textrm{dec}, \widetilde{\cD}) = \frac{1}{\widetilde{N}_d}\sum_{n=1}^{\widetilde{N}_d} \mathsf{NMSE}\Big[\widetilde{\bH}_n, \widetilde{f}_\textrm{dec}\big(\widetilde{f}_\textrm{enc}(\widetilde{\bH}_n)\big)\Big],
        \end{align}
        where $\widetilde{\cD}$ represents a dataset composed of $\widetilde{D}$ CSI matrices drawn from the digital twin distribution $\widetilde{\cH}$. It is worth noting that \eqref{eq:bound} and \eqref{eq:ml_loss2} support the intuition: When the data from the digital twin closely matches the real-world data, a DL model trained solely on the digital twin data can perform effectively on real-world data.

    \subsection{Model Refinement}
        In practical deployments, constructing a digital twin that exactly matches the real-world scenario is challenging. Imperfections in the EM 3D model, ray tracing, and hardware modeling can cause the synthetic data distribution to deviate from the real-world data distribution. In other words, these impairments in the digital twin lead to a greater discrepancy distance $\mathrm{disc}(\cH, \widetilde{\cH})$. To reduce this impact, a small set of real-world data can be utilized to fine-tune the DL model that was originally trained on digital twin data. Let $\cD_r$ represent a limited real-world dataset used to improve the DL model previously trained on $\widetilde{\cD}$. In the following, we investigate two approaches for model refinement.

        \textbf{Naive fine-tuning.} A straightforward method to enhance the DL model using the real-world dataset $\cD_r$ is through naive fine-tuning. Once the DL model has been pre-trained on the digital twin dataset $\widetilde{\cD}$, this method fine-tunes the model by optimizing the loss function in \eqref{eq:ml_loss} on the dataset $\cD_r$.

        \textbf{Rehearsal.} Naive fine-tuning can lead to overfitting when the refining dataset is limited in size. Additionally, as the DL model adapts to the refining dataset, it risks losing the generalizability learned from the digital twin, resulting in the catastrophic forgetting effect \cite{Robins1995}. To address these issues, we utilize the rehearsal approach during the model fine-tuning \cite{Robins1995}. Rehearsal involves revisiting previously learned data samples while incorporating new training data samples. By including these earlier samples in the refinement process, rehearsal strengthens or preserves previously learned information, preventing disruption from new data samples. Furthermore, it expands the effective size of the refining dataset, thereby mitigating overfitting. The rehearsal process enhances the DL model by optimizing the loss function in \eqref{eq:ml_loss} using the combined datasets $\widetilde{\cD}\cup \cD_r$.

        \textbf{System operations.} For model refinement, we assume the system operates in three phases:
        \begin{enumerate}[(i)]
            \item \textbf{Offline pre-training.} Initially, the digital twin is built on the infrastructure side. The digital twin generates data used to train the DL CSI feedback model, as introduced in \sref{sec:Direct Generalization}. Then, the trained DL models, consisting of the encoder and decoder, are transmitted to the UEs.
            
            \item \textbf{Online refinement data collection.} Once the DL model is pre-trained, the BS and UE begin employing it for CSI compression and reconstruction. Additionally, the UE has the option to send a limited set of full CSI matrices back to the BS to support model refinement.
            
            \item \textbf{Model refinement and update.} The BS collects the CSI matrices transmitted by the UEs. When enough CSI matrices have been collected, the BS can create the refinement dataset $\cD_r$ and use it to enhance the DL model. After that, the updated model is distributed to the UEs and used for the CSI compression and recovery tasks. In addition to addressing inaccuracies in the digital twin data, the system can utilize the refinement data collection and model update processes to maintain the DL model’s performance, for instance, when the environment changes.
        \end{enumerate}
        \begin{figure}[t]
            \centering
            \includegraphics[width=1\linewidth]{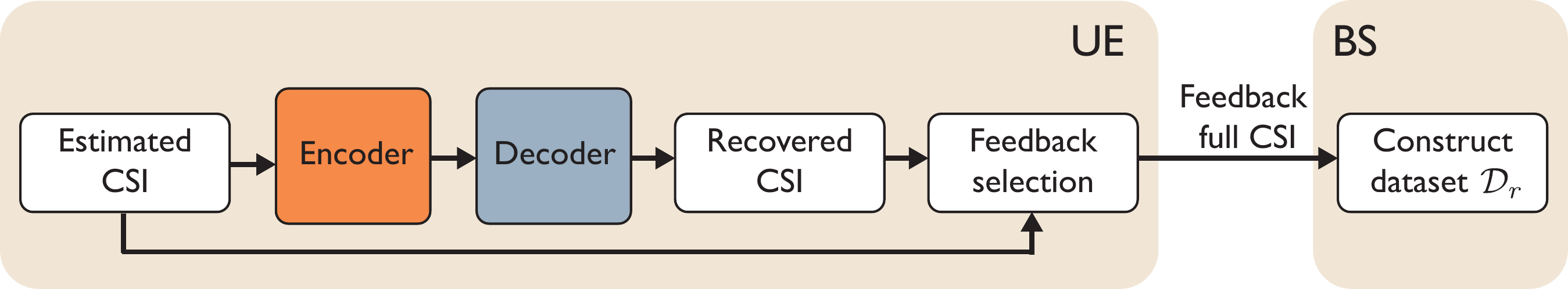}
            \caption{This figure shows the data selection process for the model refinement. The UE first compresses and recovers the CSI matrix using the DL model. The UE then calculates the reconstruction NMSE and feeds back the CSI matrices for which the NMSE is larger than a threshold $\eta$.}
            \label{fig:data_selection}
        \end{figure}
    
    \subsection{Refinement Data Selection} \label{sec:Refinement Data Selection}
        During the online data collection phase, the UE can selectively send full CSI matrices back to the BS, and this process can be optimized to improve the model refinement performance. As a baseline strategy, the UE may randomly select a subset of CSI matrices to send. However, this random selection approach can be inefficient. If the digital twin data already closely resembles the real-world data, the randomly chosen CSI matrices would strongly correlate with the digital twin data. Consequently, these collected data points offer little new insight into the real-world channel distribution, making them less useful for refining the model. To address this, we propose a heuristic approach for the UE to more effectively select CSI matrices for feedback. As illustrated in \figref{fig:data_selection}, the UE first estimates the CSI matrix and uses the encoder and decoder DL models to compress and recover the CSI matrix. Next, the UE computes the reconstruction NMSE between the original CSI matrix $\bH$ and the reconstructed CSI matrix $f_\mathrm{dec}(f_\mathrm{enc}(\bH))$ by applying \eqref{eq:nmse}. The UE then feeds back the CSI matrix whose NMSE is greater than a threshold $\eta$.

    \subsection{Deep Learning Model Design} \label{sec:Deep Learning Model Design}
        \textbf{Data Pre-Processing.} We adopt the standard approach of converting the frequency-antenna domain channel $\bH$ into the delay-angular domain to achieve a sparse representation \cite{Guo2020}. The delay-angular domain channel matrix $\bG$ is derived as
        \begin{equation}
            \bG = \bF_\textrm{d} \bH \bF^\textrm{H}_\textrm{a},
        \end{equation}
        where $\bF_\textrm{d}$ and $\bF_\textrm{a}$ are DFT matrices of dimensions $K$ and $N_t$, respectively. In the delay domain, only the first few rows of $\bG$ contain non-zero elements due to the limited delay spread. Consequently, $\bG$ is trimmed to $\bG_\textrm{trunc}$ by keeping the first 32 rows and discarding the rest. Finally, we normalize $\bG_\textrm{trunc}$ to obtain $\bG_\textrm{trunc}$ with a unit Frobenius norm:
        \begin{equation}\label{eq:norm}
            \bG_\textrm{norm} = \frac{\bG_\textrm{trunc}}{\|\bG_\textrm{trunc}\|_F}.
        \end{equation}
        Both the input and ground-truth output data for the DL model are formatted as the normalized delay-angular CSI matrix $\bG_\textrm{norm}$. Notably, this normalization process has no impact on the downlink precoding design at the BS.
        
        \textbf{NN Architecture.} We implement a NN architecture based on CSINet+ \cite{Guo2020}, with the following modifications. (i) We use the tanh activation function to replace the sigmoid activation function before the RefineNet Blocks. This activation function is intended to provide an initial estimate of the $\bG_\textrm{norm}$. The elements of the normalized CSI matrix $\bG_\textrm{norm}$ lies in $[-1,1]$ due to the normalization in \eqref{eq:norm}. Accordingly, we use the tanh activation function to ensure the values of the initial estimate align with this range. (ii) Following the final layer of the decoder in CSINet+, we added a normalization layer, similar to \eqref{eq:norm}, which adjusts the Frobenius norm of the recovered angular-delay domain CSI matrix. We observed that adding these input and output normalization layers enhance the stability and speed of the DL model’s training process. Unless stated otherwise, we use a CSI compression ratio of 1/64, where the compression ratio is the proportion of elements in the compressed CSI relative to those in the original CSI matrix.

    \subsection{Discussion on the practicality and potential overhead}
        In this subsection, we discuss the practicality and potential overhead of our proposed digital twin aided CSI feedback approach. The digital twin construction process, which involves collecting point cloud data, constructing the 3D geometry model, and modeling EM material and hardware properties, can be performed offline. Notably, building the digital twin is essentially a one-time effort for a given site. Once constructed, this digital twin can be reused for various wireless communication tasks. Also, the digital twin can be updated to reflect changes in the environment, such as new buildings or changes in foliage. This is because the digital twin has been modularized into four components: 3D geometry model, EM material model, ray tracing simulator, and hardware model. This modularity allows for straightforward updates to the digital twin without requiring a complete reconstruction. Following construction, synthetic data can be generated from the digital twin, and the DL model can be pre-trained on this synthetic data. Both the synthetic data generation and the DL model pre-training steps can be performed offline, which significantly reduces the overhead typically associated with collecting extensive real-world CSI data.

        Next, we discuss the practicality and potential overhead introduced by the model refinement process, which includes three main components:
        \begin{itemize}
            \item \textbf{Feedback overhead (sending selected full CSI):} When the UE identifies a CSI sample that indicates poor compression performance with the current model, it sends this full CSI matrix back to the BS for model refinement. This process naturally introduces communication overhead. To quantify this, we can consider the data size. For a 32-antenna array and 256 subcarriers, the channel is a 32×256 complex-valued matrix, which contains 8192 elements. If we use single-precision floating-point format (float32), each complex element requires 8 bytes (4 for real, 4 for imaginary). Thus, the total data size for one CSI sample would be 8192 elements × 8 bytes/element = 65536 bytes = 64 KB, which is a relatively small amount of data. Notably, unlike regular CSI feedback for instantaneous downlink transmission, there is no strict real-time requirement for sending back this selected full CSI. This allows for more flexible scheduling of the feedback, potentially during periods of lower network load, which further mitigates its impact.
            \item \textbf{Computation overhead (retraining/updating models at the BS):} Next, once the BS has collected a sufficient amount of data, it initiates the model refinement process, which incurs computational overhead. It is common practice in the literature to employ relatively compact models for CSI compression, considering the limited computational capacity of user devices. For instance, the CSINet+ model adopted in our work has an encoder parameter size of 0.26 MB and a decoder parameter size of 0.36 MB. Also, since the model has learned generalizable patterns from the digital twin data, both the number of real-world data points and the number of iterations required during the refinement phase can be significantly reduced.
            \item \textbf{Communication overhead (distributing the retrained model to the UE):} Finally, after refinement, the BS must distribute the updated model to the UEs, leading to communication overhead. As mentioned, CSI compression models typically have small sizes, which minimizes the communication burden during model distribution. Moreover, ongoing research is exploring ways to further reduce this overhead. For example, prior work has investigated model quantization~\cite{Lu2021}, model pruning~\cite{Singh2024}, and knowledge distillation~\cite{Cui2024} for the CSI compression.
        \end{itemize}

\section{Simulation Setup} \label{sec:Simulation Setup}
    In this section, we explain the simulation setup, including the target scenario, the digital twin scenario, and the baseline scenarios. We also introduce the dataset generation process.

    \subsection{Scenario Setup}
    In our evaluation, we utilize three distinct scenario types: (i) the target scenario, (ii) the digital twin scenario, and (iii) the baseline scenario. 

\begin{figure}[t]
        \centering
        \includegraphics[width=0.9\linewidth]{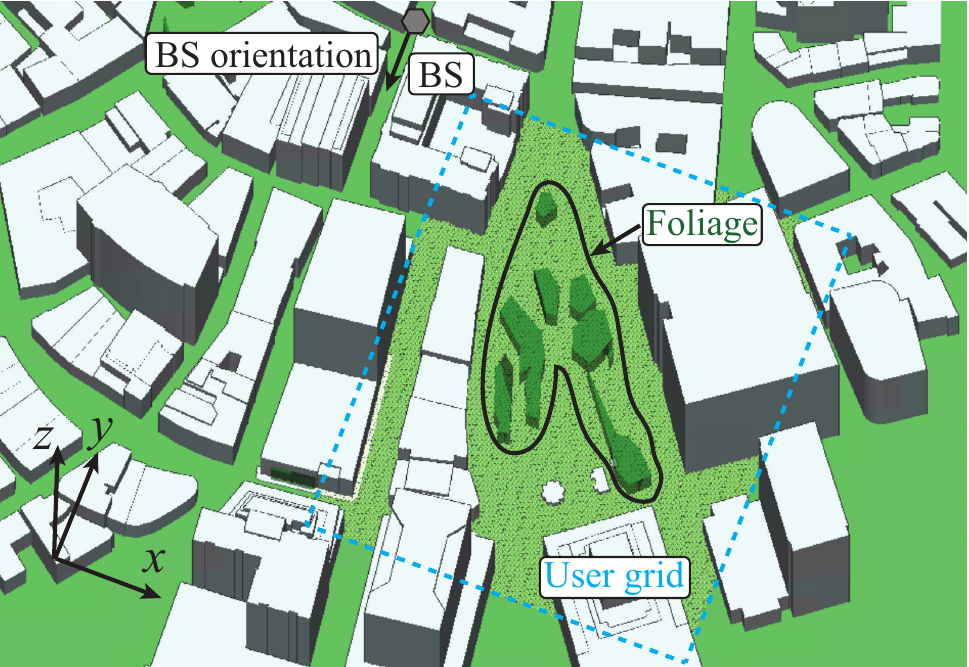}
        \caption{This figure illustrates the geometric layout of the target scenario, representing a real-world section of Downtown Boston. The BS is positioned along a vertical street, oriented toward the negative y-axis. The service area is highlighted in blue, with a foliage object situated in the middle.} 
        \label{fig:scenario_boston}
    \end{figure}

    \textbf{Target scenario.} The target scenario serves as the "real-world" setting in the evaluation. \figref{fig:scenario_boston} provides the bird's-eye view of this scenario, depicting an actual urban area in Boston. The target scenario includes a BS, a service area, and various building and foliage objects. The BS is equipped a 32-antenna uniform linear array (ULA) at a height of 15 meters, oriented towards the negative y-axis. The service area (annotated in blue) covers where the UE can be located. This area spans 200 meters by 230 meters, with the UE positioned at a height of 2 meters. The service area is divided into a grid of multiple UE locations, spaced 0.37 meters apart.

    \begin{figure}[t]
        \centering
        \includegraphics[width=0.9\linewidth]{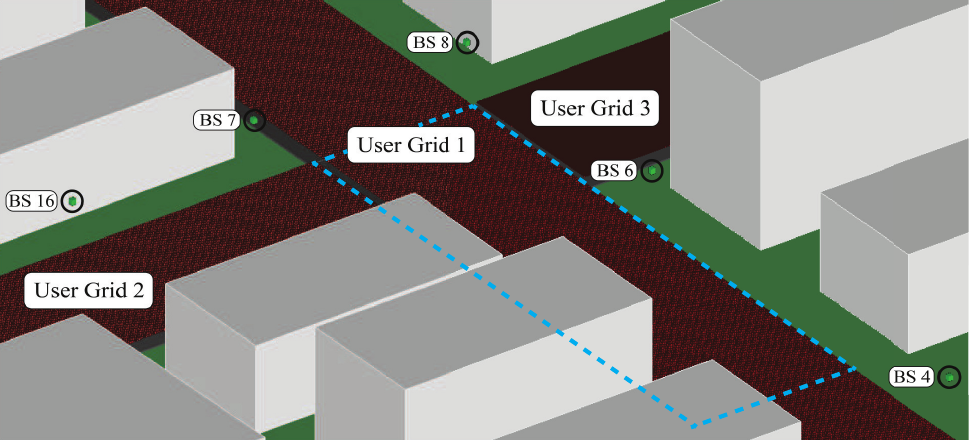}
        \caption{This figure illustrates the geometric layout of the baseline scenario, which is the O1 scenario in the DeepMIMO dataset \cite{deepmimo}. We choose 'BS 6' as the base station and select rows 1150 to 1650 as the service area, which is marked in blue.}
        \label{fig:scenario_o1}
    \end{figure}

    \textbf{Digital twin scenario.} Based on the geometry of the target scenario, we construct the digital twin as introduced in~\sref{sec:Digital Twin Construction}. For 3D mesh processing, we use MeshLab~\cite{meshlab}. We import the original building models from the target scenario into MeshLab and apply Poisson disk sampling~\cite{Corsini2012} to create a point cloud. The sampling density adjusts the data collection overhead required to build the 3D geometry model. Next, we estimate the normal vectors of the sampled points and reconstruct the 3D geometry model with screened Poisson surface reconstruction~\cite{Kazhdan2013}. The reconstructed models typically contain more faces than the original models, according to empirical observations. To align the ray-tracing complexity with the target scenario, we reduce the number of faces to match the original building models and employ quadric edge collapse decimation~\cite{Farland1997} for this purpose. Thus, the geometry fidelity of the digital twin depends on the data collection overhead, the mesh reconstruction process, and the face reduction method. Furthermore, we exclude foliage objects from the digital twin scenario because seasonal changes and random growth patterns complicate accurate modeling in practice. The digital twin scenario incorporates the same BS and service area as the target scenario.

    \textbf{Baseline Scenario.} To highlight the effectiveness of the site-specific digital twin CSI data, we are interested in a comparative analysis with CSI data generated from non-site-specific scenarios. Accordingly, we evaluate the following two baseline scenarios:
        \begin{enumerate}[(i)]
            \item \textbf{DeepMIMO O1.} We employ the O1 scenario from the DeepMIMO dataset \cite{deepmimo}, which can be seen as uncorrelated site-specific digital twin. As shown in \figref{fig:scenario_o1}, we select the ``BS 6" as the BS and choose rows 1150 to 1650 of ``User Grid 1" as the discretized service area (annotated in blue).
            \item \textbf{WINNER II.} We adopt the ``Urban Macro Cell" scenario of the WINNER II channel model \cite{WinnerII} as the second baseline scenario to compare our proposed approaches with a generic statistical dataset. The WINNER II channel model is a widely used channel model for evaluating wireless communication systems.
        \end{enumerate}   

    \subsection{Dataset Generation} \label{sec:Dataset Generation}
        We employ Wireless Insite~\cite{Remcom} to perform accurate 3D ray tracing between the BS and all UE positions in the target, digital twin, and baseline scenarios. The ray tracing searches the propagation paths between the BS and each UE position, according to the geometry layout. For every propagation path, it generates path parameters, including complex gain $\alpha_l$, propagation delay $\tau_l$, azimuth AoD $\phi_l$, and elevation AoD $\theta_l$. The ray tracing angle resolution is set to 0.25 degrees. For the propagation model, each path can undergo a maximum of 4 reflections before reaching the UE. We assume that the BS antenna array has a 180-degree FoV. Also, the BS serves the UE at the 3.5 GHz band with 256 subcarriers and $30$ kHz spacing. For the EM materials, we consider the building material as concrete and the terrain material as wet earth, following the ITU default parameters at 3.5 GHz. The foliage objects are modeled as an attenuation material, where the attenuation coefficient is set to 1 dB/m. Using the path parameters, we generate the CSI matrices $\bH$ between the BS and UE positions by applying \eqref{eq:delayd_channel} and \eqref{eq:frequency_channel} with the DeepMIMO channel generator \cite{deepmimo}. Lastly, we perform the pre-processing mentioned in \sref{sec:Deep Learning Model Design} to transform $\bH$ into the delay-angular domain representation $\bG_\textrm{norm}$. 

\section{Evaluation Results} \label{sec:Evaluation Results}
    In this section, we evaluate the performance of the digital twin aided CSI feedback. First, we analyze the benefits of using digital twins for CSI reconstruction. Next, we examine the effectiveness of the proposed data selection and model refinement approaches. In addition, we explore how the fidelity of the digital twin affects the performance. Finally, we discuss the practical implications of digital twins by comparing the effects of digital twinning and channel estimation errors.

    \begin{figure}[t]
            \centering
            \includegraphics[width=0.985\linewidth]{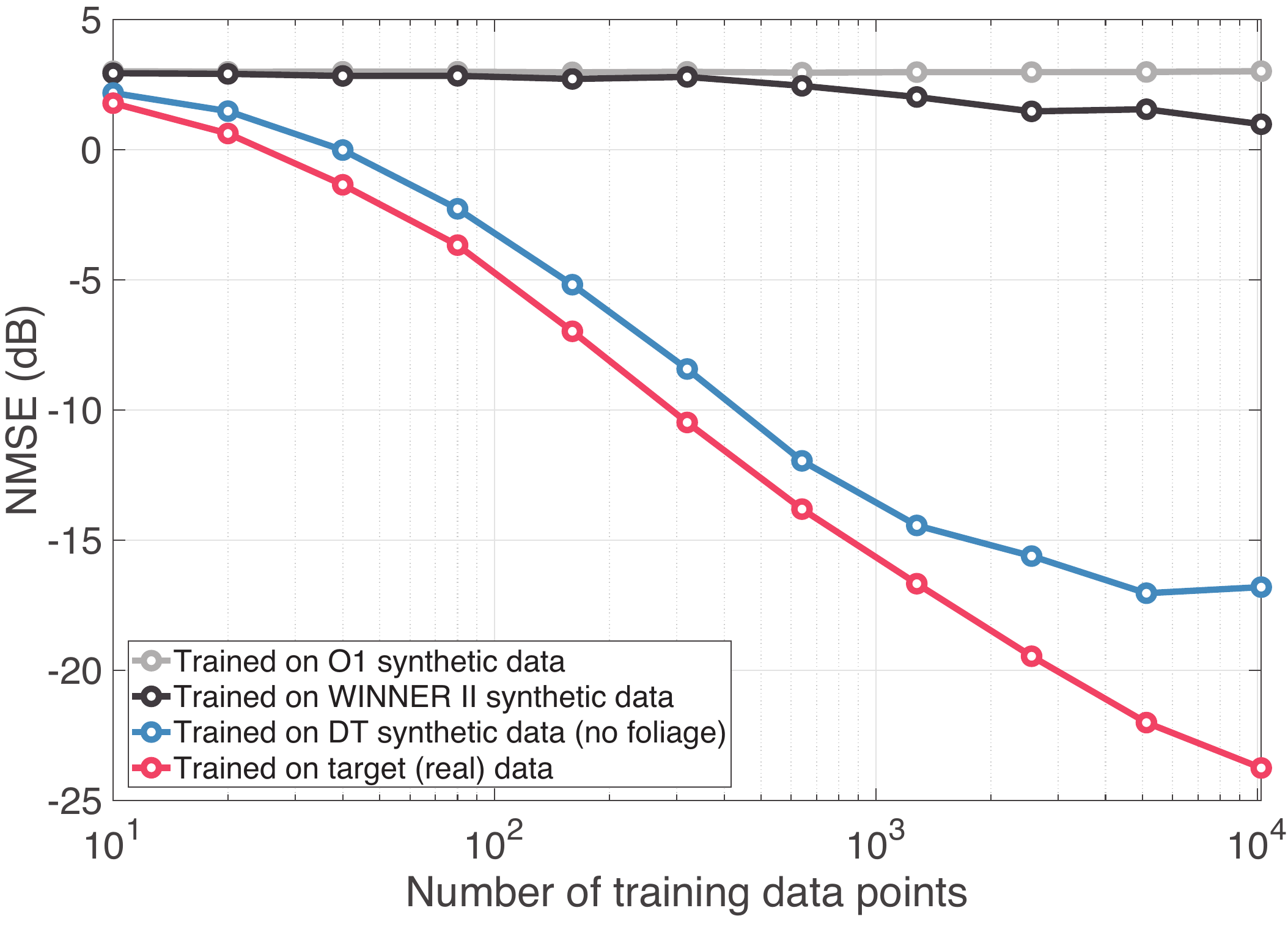}
            \caption{This figure presents the NMSE performance of the direct generalization approach. All NMSE evaluations are conducted on target data that was not part of the training set. For comparison, the performance of the DL model trained on the data from the target and the baseline scenarios are also presented.}
            \label{fig:data_size}
    \end{figure}

    \begin{figure}[t]
        \centering
        \includegraphics[width=1\linewidth]{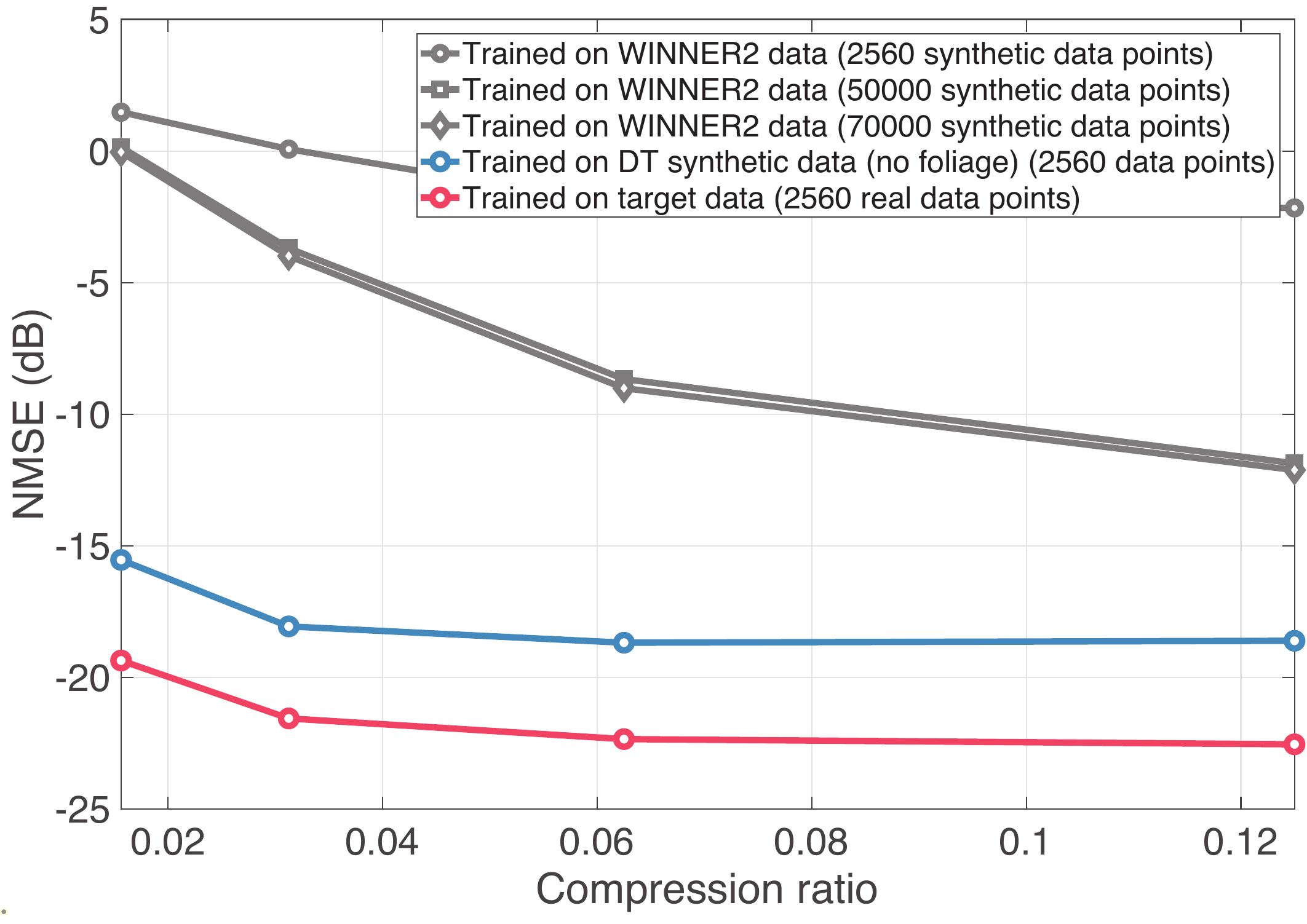}
        \caption{This figure presents the NMSE performance across different compression ratios. The compression ratio is set to $\{1/64, 1/32, 1/16, 1/8\}$. Adjusting the compression ratio allows for a trade-off between feedback overhead and CSI reconstruction performance.}
        \label{fig:compression_ratio}
    \end{figure}

    \subsection{Can digital twins help CSI reconstruction?}
        In this subsection, we explore how digital twins contribute to enhancing CSI reconstruction. Specifically, we evaluate the performance of a DL model trained on site-specific CSI data generated by a high-fidelity digital twin, where the only difference from the target scenario is the presence of foliage objects. To demonstrate the advantages of a site-specific digital twin, we first analyze the performance of DL models trained on baseline scenarios. \figref{fig:data_size} presents the CSI reconstruction NMSE when the DL model is trained on different amounts of synthetic data and tested on the target dataset. The results show that models trained on baseline scenarios do not generalize well to the target scenario. Moreover, increasing the number of training data points provides only marginal improvements in NMSE performance. This limitation arises because the DeepMIMO O1 scenario has a significantly different geometric layout compared to the target scenario. Additionally, the WINNER II scenario is a generic statistical dataset that does not necessarily capture the site-specific characteristics of the target environment. Thus, the substantial differences in CSI distributions between these scenarios hinder the generalization ability of the DL model for CSI reconstruction. In contrast, the EM 3D model in the site-specific digital twin enables it to generate CSI data that closely approximate the target data. Therefore, the DL model trained on the digital twin data can generalize well to the target scenario. Notably, this approach eliminates the need for real-world CSI data collection since it does not rely on any target data during training. These findings highlight the critical role of site-specific digital twins in effectively training DL models for CSI reconstruction.

        Next, we evaluate the performance of CSI reconstruction under varying compression ratios. The compression ratio can be adjusted to balance feedback overhead and the performance of the DL model. A lower compression ratio results in greater information loss during the CSI compression process, leading to poorer NMSE performance. \figref{fig:compression_ratio} compares the site-specific data generated by the digital twin with the generic WINNER II dataset. The results demonstrate that the proposed approach significantly outperforms the WINNER II scenario, and increasing the number of training data points in the WINNER II dataset provides only limited NMSE improvements. Interestingly, the WINNER II scenario shows more noticeable performance gains at higher compression ratios. This is because the WINNER II dataset is a generic dataset and requires a larger latent space to capture site-specific characteristics that are not inherently present in its data. In contrast, the site-specific digital twin dataset inherently contains the critical channel characteristics of the target scenario, making it more efficient for CSI reconstruction. Since the digital twin only contains the relevant site-specific information, it requires less feedback overhead to achieve the same or even better performance than a generic dataset. This further reinforces the effectiveness of site-specific digital twins, as they not only improve CSI reconstruction accuracy but also reduce the CSI feedback burden by eliminating unnecessary non-site-specific information.

        \begin{figure}[t]
            \centering
            \includegraphics[width=0.995\linewidth]{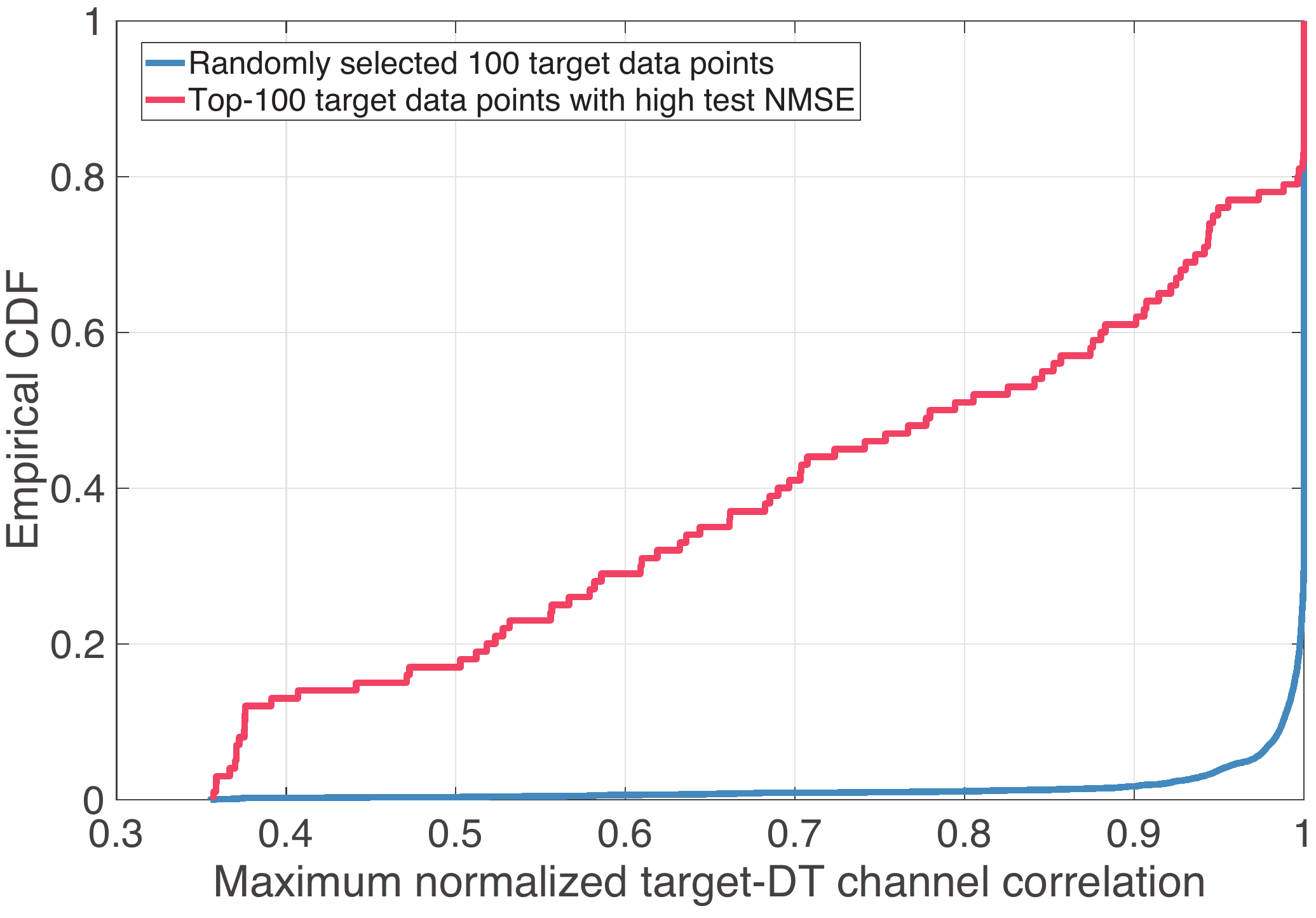}
            \caption{This figure shows the empirical CDF of the normalized correlation between the target and digital twin CSI matrices. Compared to randomly selected CSI matrices, the proposed data selection method is effective in selecting the most informative CSI matrices for model refinement.}
            \label{fig:correlation}
        \end{figure}
        \begin{figure}[t]
            \centering
            \includegraphics[width=1\linewidth]{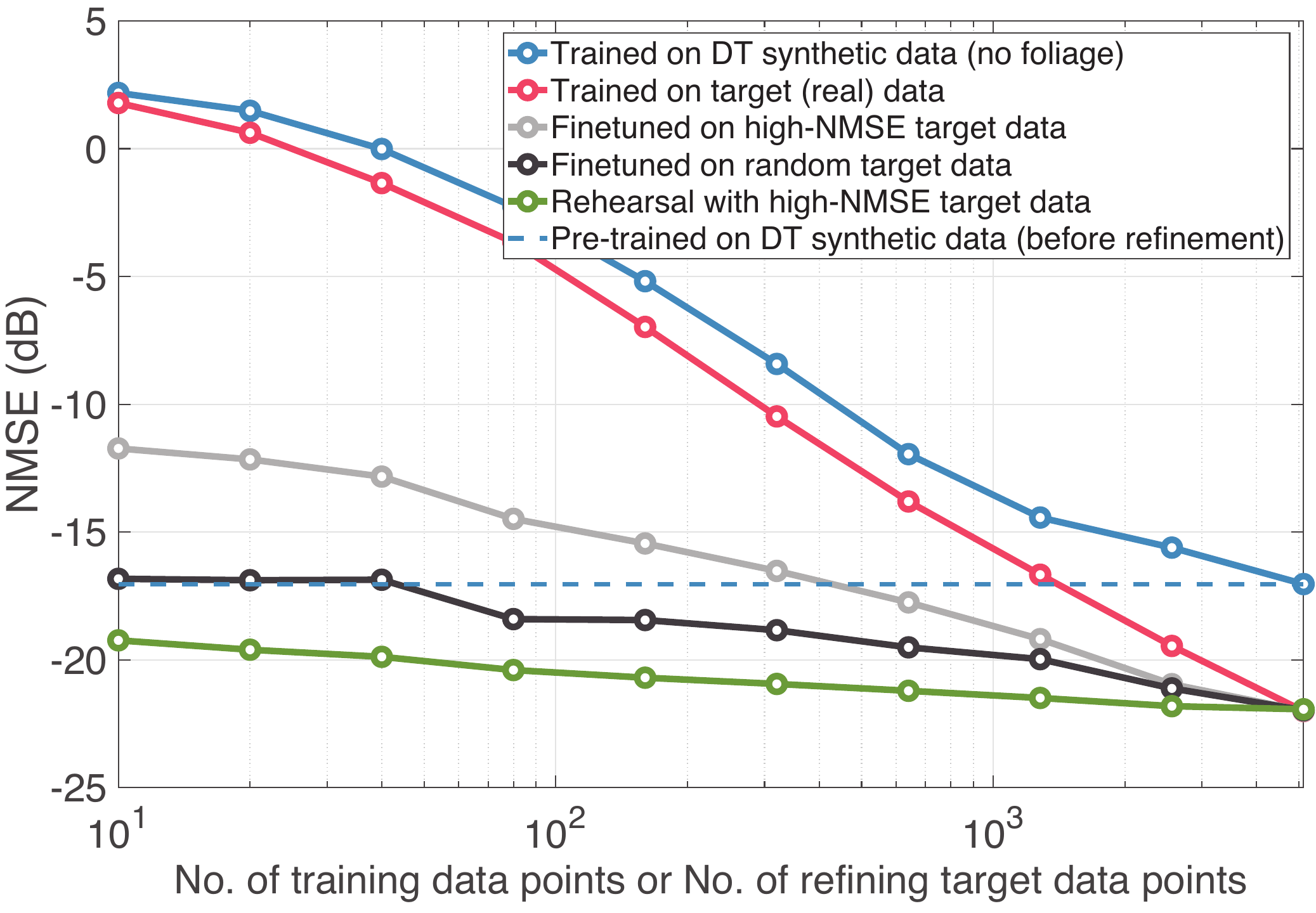}
            \caption{This figure shows the NMSE performance of the direct generalization and three model refinement approaches. All NMSE performance is evaluated on the target data unseen in the training and refining. Among the three model refinement approaches, the rehearsal approach achieves the best performance.}
            \label{fig:data_size_refine}
        \end{figure}
        \begin{figure}[t]
            \centering
            \includegraphics[width=1\linewidth]{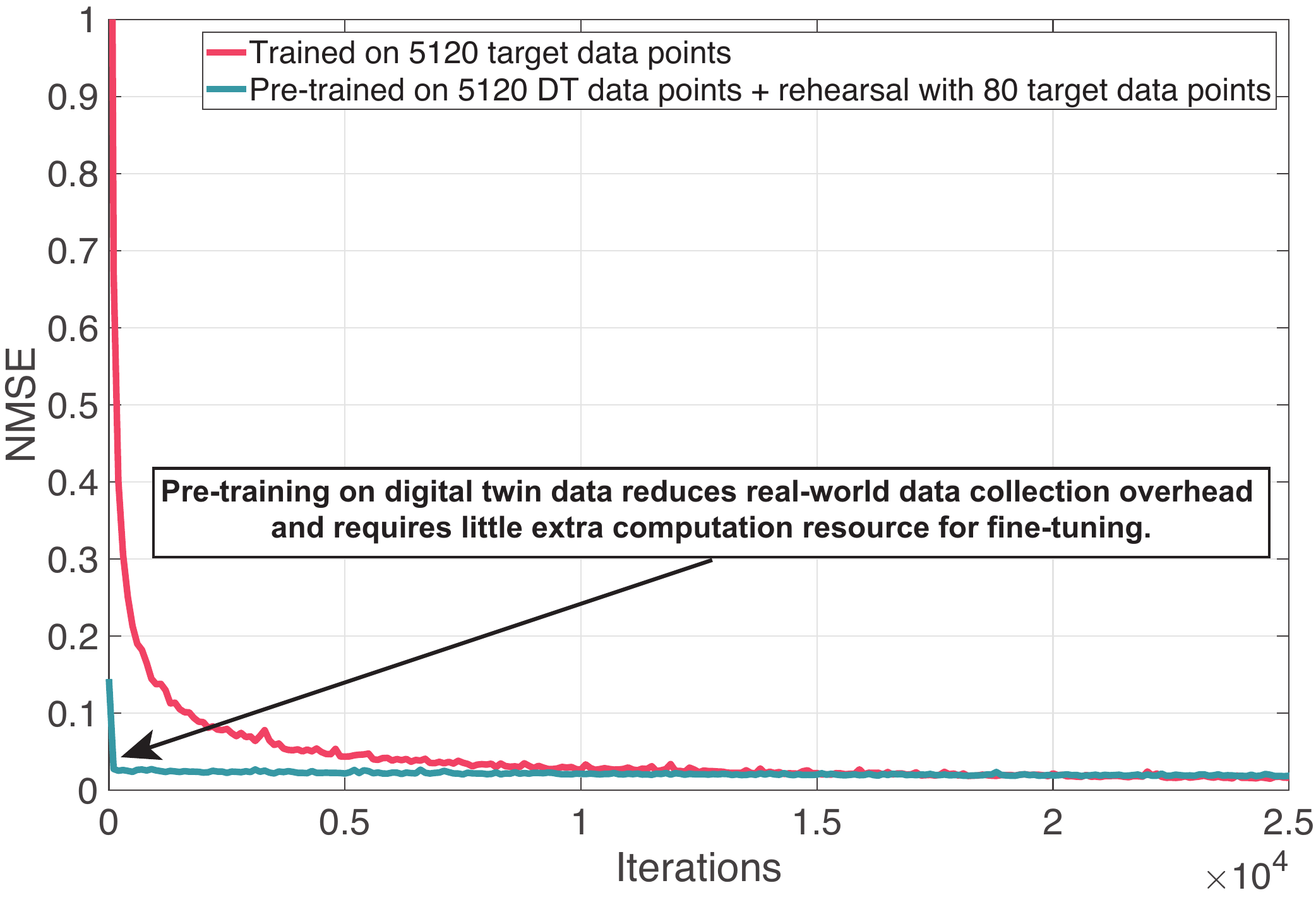}
            \caption{This figure presents the NMSE performance of CSI reconstruction evaluated across varying numbers of fine-tuning iterations. For comparison, we also show the performance of the DL model trained solely on the target data. Notably, to achieve comparable performance, pre-training on digital twin data requires significantly fewer data points and minimal fine-tuning iterations.}
            \label{fig:nmse_vs_iter}
        \end{figure}
        
        \begin{figure*}[!t]
            \centering

            \begin{subfigure}[t]{0.31\textwidth}
                \centering
                \includegraphics[width=\textwidth]{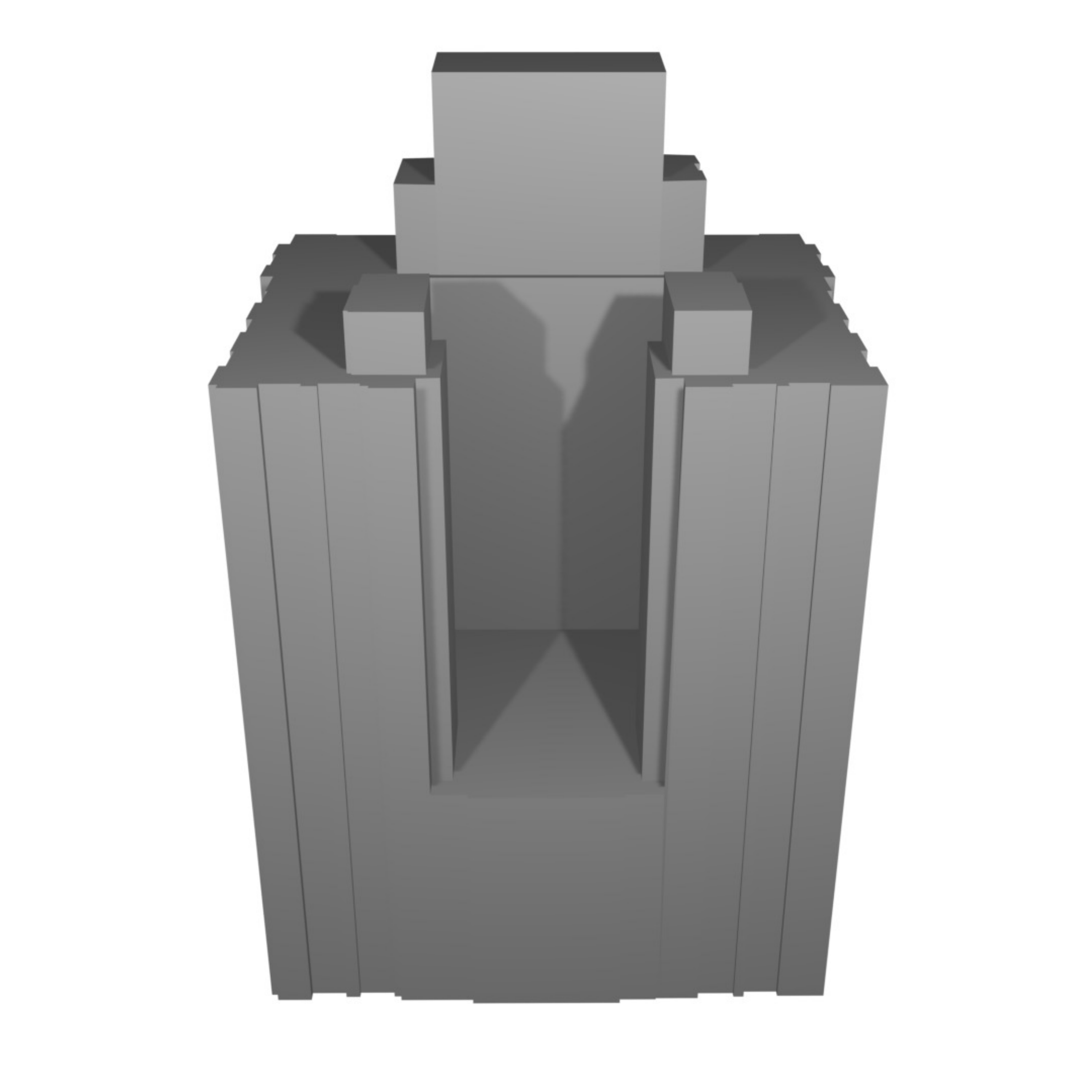}
                \caption{Original building model}
            \end{subfigure}
            \hfill
            \begin{subfigure}[t]{0.31\textwidth}
                \centering
                \includegraphics[width=\textwidth]{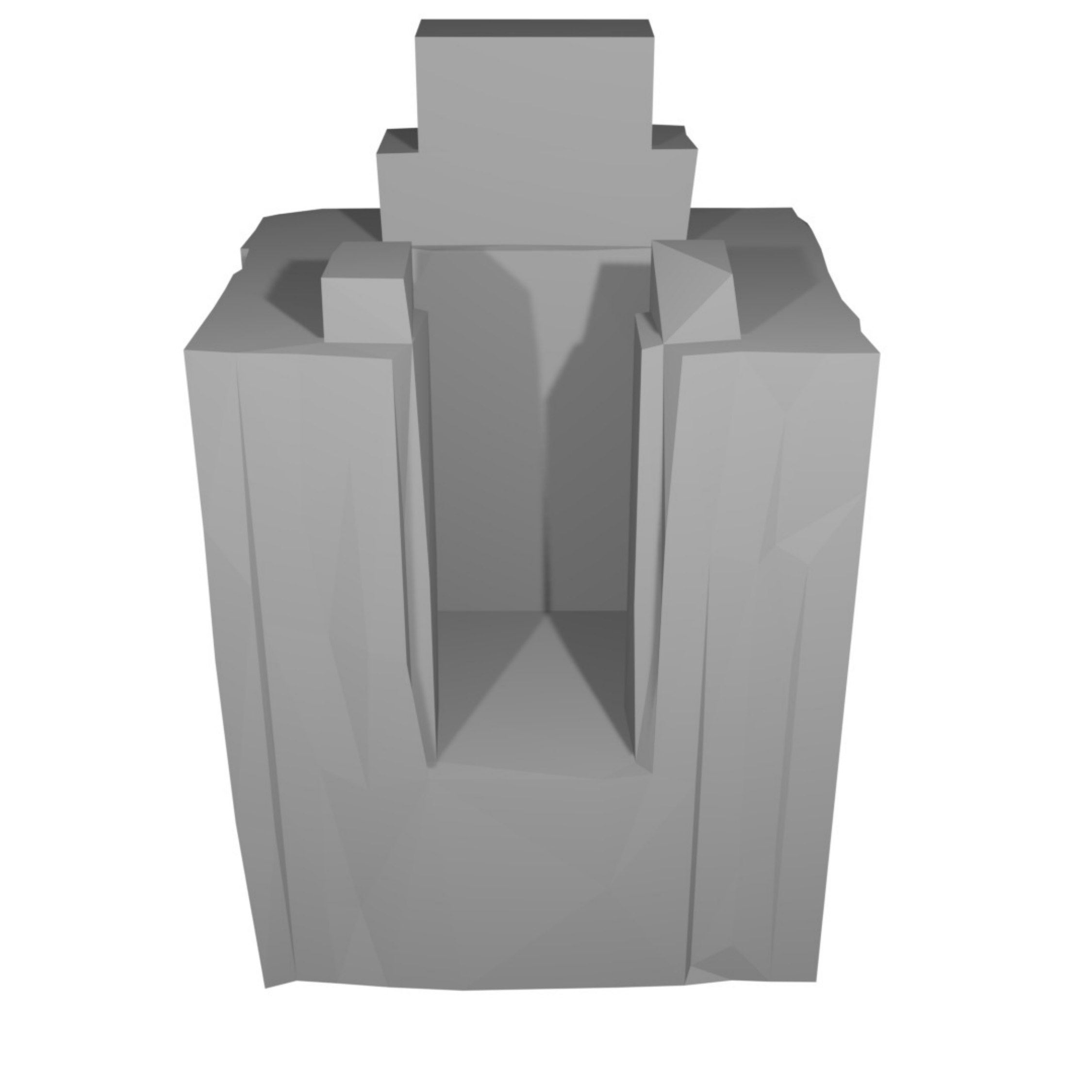}
                \caption{Sampling density = 2 points/$\textrm{m}^2$ \\ F1-score = 99.71}
            \end{subfigure}
            \hfill
            \begin{subfigure}[t]{0.31\textwidth}
                \centering
                \includegraphics[width=\textwidth]{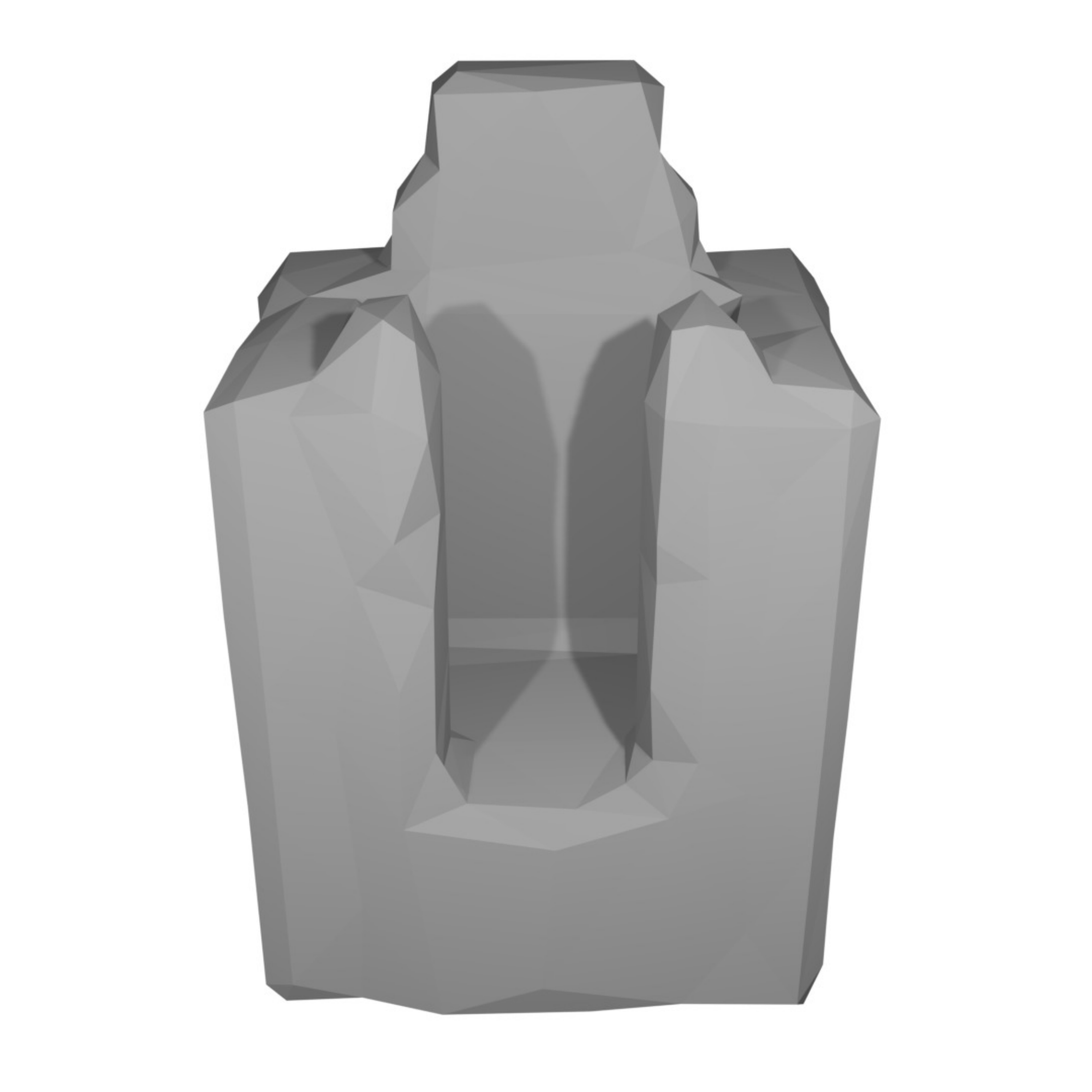}
                \caption{Sampling density = 0.05 points/$\textrm{m}^2$ \\ F1-score = 97.19}
            \end{subfigure}
            
            \caption{An example of the original building model and the reconstructed building model with different sampling densities and the corresponding F1-scores. A higher F1-score indicates a higher similarity between the original and reconstructed building models.}	
            \label{fig:building_comparison}
        \end{figure*}
        \begin{figure}[t]
            \centering
            \includegraphics[width=0.98\linewidth]{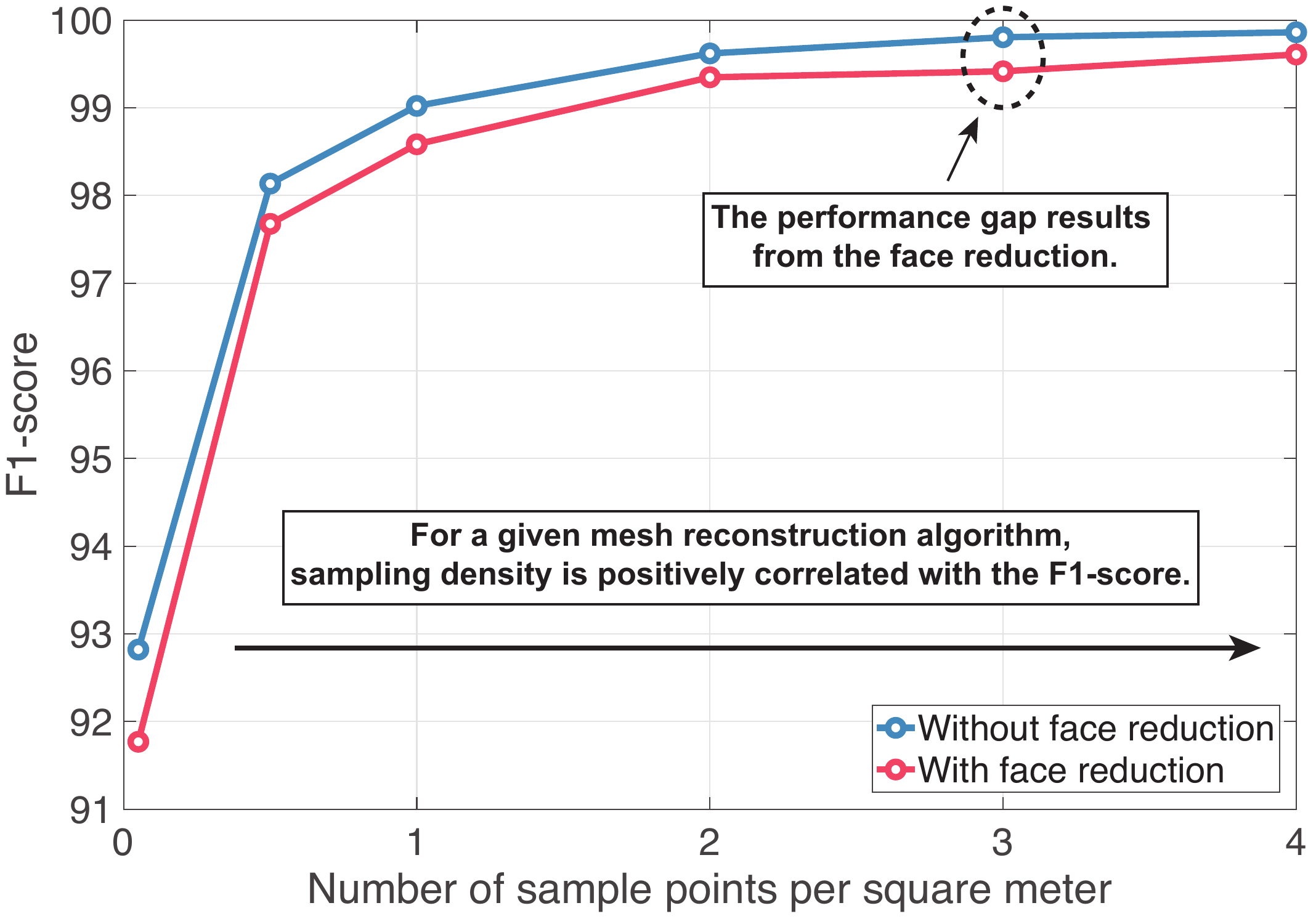}
            \caption{This figure shows the F1-score of reconstructed building models at different sampling densities. The average F1-score for each density level is computed across all buildings in the scene. Also, the performance gap introduced by the face reduction algorithm is presented.}
            \label{fig:f1_score}
        \end{figure}
        \begin{figure}[t]
            \centering
            \includegraphics[width=0.85\linewidth]{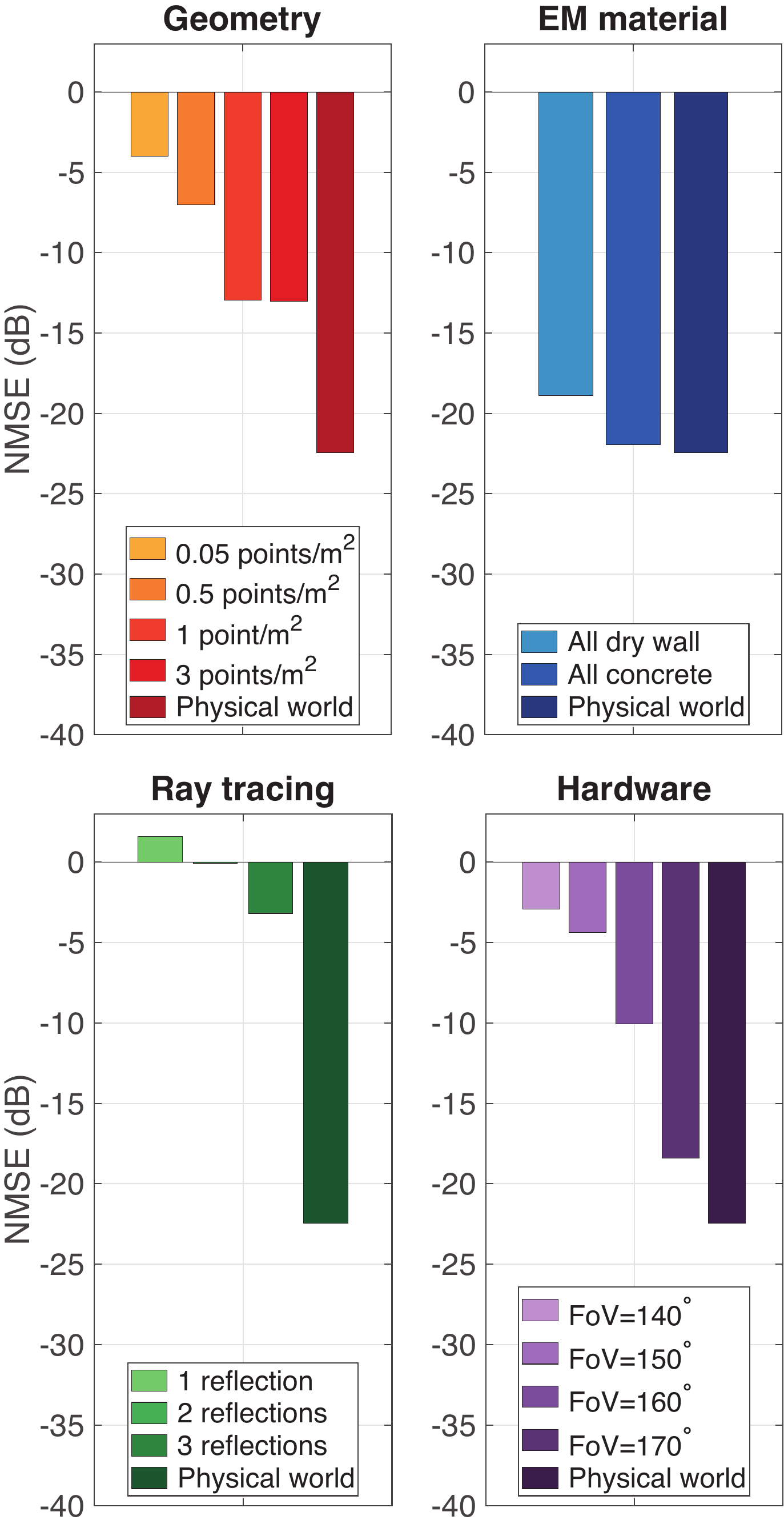}
            \caption{This figure presents the sensitivity analysis of the DL model performance with respect to 3D geometry, EM material, ray tracing, and hardware modeling. The results indicate that, in CSI reconstruction, 3D geometry, ray tracing, and hardware modeling have a greater impact on performance compared to EM material.}
            \label{fig:sensitivity}
        \end{figure}
        \begin{figure}[t]
            \centering
            \includegraphics[width=1\linewidth]{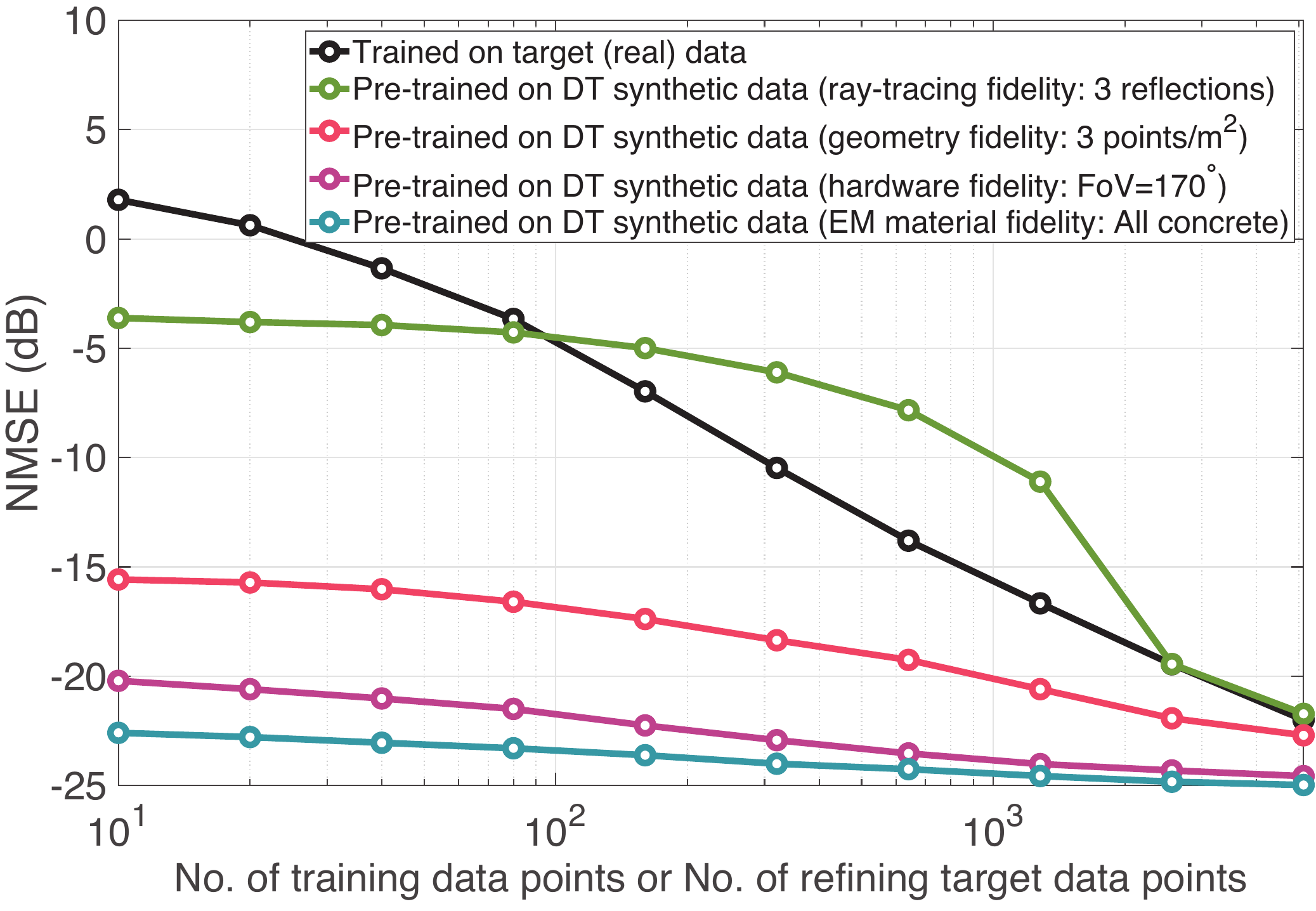}
            \caption{This figure presents the refinement performance of a DL model pre-trained on digital twin data where one of the four fidelity aspects is impaired. Better initial performance from the pre-trained model leads to more effective refinement, particularly in mitigating real-world data collection overhead.}
            \label{fig:sensitivity_finetune}
        \end{figure}

    \subsection{Can the DL model trained on the digital twin data be further refined?}
        In the previous subsection, we have shown that the DL model trained on the digital twin data performs well on the target scenario. However, a performance gap remains due to the mismatch between the digital twin and the target scenarios. To address this, we investigate whether the DL model can be further refined using a small amount of target data to enhance its performance. In this subsection, we evaluate the effectiveness of the proposed CSI online data selection and model refinement strategies. First, during the offline pre-training phase, we train a DL model with 5120 data points produced by the digital twin. This pre-trained model is then used to select the target CSI data for subsequent fine-tuning. Specifically, it compresses and reconstructs all the CSI matrices in the target training dataset, and the reconstruction NMSE for each of matrix is calculated. To assess the proposed data selection approach, we calculate the highest normalized correlation between each target CSI matrix and the full set of digital twin CSI matrices. Let $\widetilde{\cD} = \{\widetilde{\bH}_1, \hdots, \widetilde{\bH}_{\widetilde{D}}\}$ denote the dataset containing the $\widetilde{D}$ digital twin CSI matrices. The maximum correlation calculated from the target CSI matrix $\widetilde{\bH}$ can be written as
        \begin{equation}
        \max_{i=1,\hdots,\widetilde{D}} \frac{|u(\bH)^\textrm{H}u(\widetilde{\bH}_i)|}{\|u(\bH)\|_{2}\|u(\widetilde{\bH}_i)\|_{2}},
        \end{equation}
        where $u(\bH)$ flattens the CSI matrix $\bH$ to one-dimensional, and $\|\cdot\|_{2}$ denote the $L_2$ vector norm. A high maximum correlation (near one) suggests that a similar CSI matrix exists in the digital twin dataset, meaning the target CSI matrix adds little new information and is less useful for fine-tuning. In \figref{fig:correlation}, we present the empirical cumulative distribution function (CDF) for the normalized correlation derived from the target CSI matrices that have the top-100 highest recovery NMSE. For benchmarking, we include a similar plot using 100 randomly select CSI matrices. The results indicate that the target CSI matrices with high reconstruction NMSE lead to a notably lower normalized correlation. Thus, these target CSI matrices more effectively capture the mismatch between the target and digital twin data, underscoring the efficacy of the proposed data selection approach.

        \figref{fig:data_size_refine} presents the NMSE performance achieved by refining the pre-trained model through three distinct approaches. For comparison, we also provide the performance of the pre-trained model without any refinement. Refining the model with a limited set of randomly chosen target data does not improve NMSE performance, as these data are unlikely to reflect the differences between the target and digital twin CSI distributions. When the model is fine-tuned on a small number of high-NMSE target data, the performance degrades compared to the pre-trained model. While the high-NMSE target data can capture discrepancies between the target and digital twin, it deviates considerably from the digital twin data employed during pre-training. As a result, directly fine-tuning the pre-trained model with these types of data can interfere with its previously learned patterns. In contrast, refining the model through rehearsal can effectively adjust for discrepancies between the digital twin and target scenarios while maintaining the patterns learned earlier. This allows the pre-trained model to be further enhanced using only a small amount of target data.

        Next, we study the computational overhead of refining the DL model. In \figref{fig:nmse_vs_iter}, we present the NMSE performance of the DL model with respect to the number of fine-tuning iterations. Specifically, we show the performance of a DL model pre-trained on 5120 digital twin data points and then refined on 80 target data points using the rehearsal approach. For comparison, the performance of a DL model trained solely on 5120 target data points is also presented. The results demonstrate that, to achieve a comparable performance level, the model pre-trained on digital twin data converges much faster, doing so within 200 iterations. This indicates that pre-training the DL model on digital twin data not only lowers the demand for real-world data but also significantly reduces the computational overhead during the model refinement phase. Essentially, pre-training with digital twin data provides a robust initialization for the DL model, enabling it to learn the important patterns of the target scenario more efficiently.

    \subsection{What is the impact of digital twin fidelity?}
        In this subsection, we examine how the fidelity of a digital twin affects CSI reconstruction. We begin by evaluating the role of point cloud data collection, the first step in constructing a digital twin. The sampling density of the point cloud directly affects the fidelity of the resulting 3D geometry model, as a denser point cloud can capture more detailed structural information of the buildings. \figref{fig:building_comparison} visualizes the reconstructed building models with different sampling densities. Models with higher sampling densities are more similar to the original model, with sharper edges and more defined contours. From these visualizations, it is evident that the geometry fidelity affects the subsequent ray-tracing process. For instance, if the face normal of the reconstructed building model is not accurate, signal interactions with building surfaces will also be misrepresented. This leads to a mismatch between the digital twin and the target scenario, which in turn impacts the performance of the DL model. To quantify the geometry fidelity, we compute the F1-score between the original and reconstructed building models, as shown in \figref{fig:f1_score}. The results indicate that the data collection overhead is positively correlated with the geometry fidelity of the digital twin. Notably, after reaching a certain sample density, the fidelity is mainly bounded by the limitations of the mesh reconstruction and face reduction algorithms. Although this study does not compare different mesh reconstruction and face reduction algorithms, they remain essential for the fidelity of the digital twin.

        Then, we investigate the impact of the digital twin fidelity on CSI reconstruction with respect to the 3D geometry, EM material, ray tracing, and hardware model. Specifically, we conduct a sensitivity analysis by adjusting the parameter in one aspect of digital twin fidelity while keeping the other aspects fixed. For the 3D geometry, we vary the sampling densities of the building models, while the other fidelity aspects follow the description in \sref{sec:Dataset Generation}. For EM material, instead of explicitly modifying permittivity and conductivity values, we categorize materials by name to represent fidelity levels. For instance, we assess different accuracy levels by assigning all objects as either concrete or drywall, without altering 3D geometry or ray tracing. In this case, drywall serves as a lower-fidelity material than concrete relative to the target scenario. For the ray tracing, we vary the maximum number of reflections from 1 to 4 while keeping the remaining fidelity aspects unchanged. Finally, for the hardware model, we vary the FoV of the digital twin from 140 degrees to 170 degrees. The results in \figref{fig:sensitivity} indicate that, in addition to 3D geometry, ray tracing and the hardware model also have a significant impact on the performance of the DL model. For ray tracing, the parameters directly influence multi-path fading and determine whether a path exists between the BS and UE. For example, reducing the maximum number of reflections from 4 to 1 leads to a significant performance drop. This occurs because, with fewer reflections, ray tracing may fail to find a path for a user that requires multiple reflections to reach the BS, resulting in a significant coverage drop and CSI distribution mismatch. Similarly, the hardware model also plays a crucial role in determining coverage and CSI distribution. For instance, reducing the FoV from 180 degrees to 140 degrees leads to a significant performance drop because a reduced FoV excludes certain paths from the channel synthesis, resulting in a distribution mismatch between the digital twin and the target scenario. In contrast, the EM material modeling has a relatively smaller impact on the performance of CSI reconstruction.
        
        Based on the sensitivity analysis, in \figref{fig:sensitivity_finetune}, we present the refinement performance of the DL model pre-trained on digital twin data with one fidelity aspect impaired. The results show a positive correlation between the initial pre-training performance (or the mismatch between the digital twin and target scenario) and the subsequent refinement performance. For instance, fidelity aspects with a larger initial impact, such as ray tracing, demonstrate only marginal gains after refinement. However, these marginal gains are still notable when compared to training the model solely on target data from scratch, especially when limited target data is available. Conversely, for fidelity aspects with a smaller initial impact, such as EM material, pre-training on digital twin data provides a significant gain in reducing real-world data collection overhead. This result further underscores the importance of conducting such sensitivity analyses. Specifically, understanding which twinning fidelity aspects a particular wireless communication task and site are most sensitive to enables more efficient digital twin development and deployment strategies.

        \begin{figure}[t]
            \centering
            \includegraphics[width=1\linewidth]{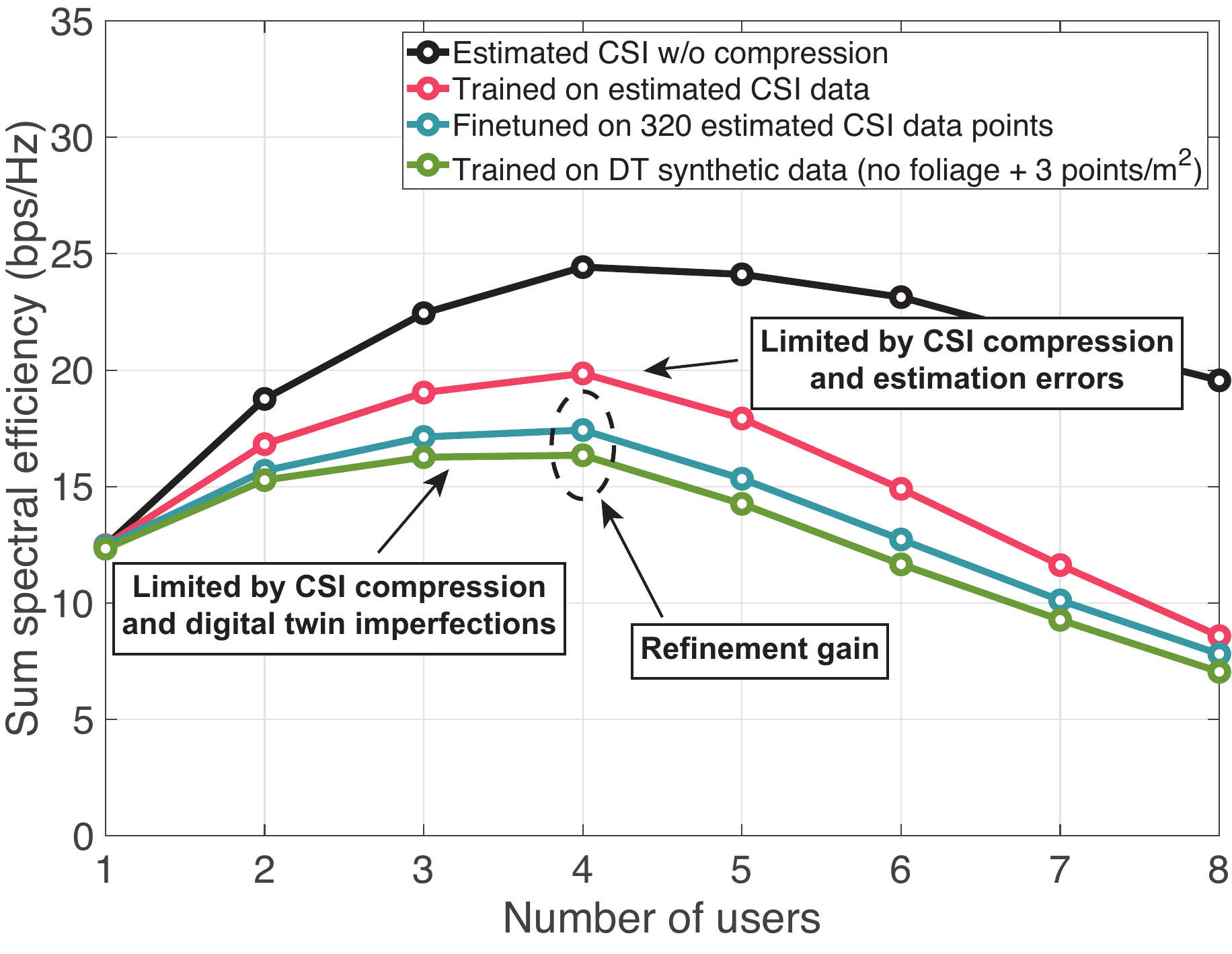}
            \caption{This figure shows the spectral efficiency performance of the proposed digital twin aided CSI feedback approach. We adopt zero forcing to compute the precoding matrix based on the reconstructed channel, and the spectral efficiency is evaluated based on the definition in \eqref{eq:sum_rate}.}
            \label{fig:spectral_efficiency}
        \end{figure}
    \subsection{Comparison of twinning and channel estimation imperfections}
        Finally, we explore the practical implications of digital twin-aided CSI feedback by comparing twinning and channel estimation imperfections. In practice, the channel obtained via channel estimation is inherently noisy due to thermal noise and active components in the receiver. Thus, this noisy channel information is actually what the encoder compresses at the UE. Interestingly, a similar form of ``imperfection'' emerges from the digital twin itself: Without explicitly modeling hardware characteristics (as is currently the case in our work), a ray tracer generates a ``clean'' channel, but its inherent modeling errors render the synthetic channel ``imperfect'' when compared to real-world channels. This parallel in data characteristics motivates an intriguing comparison between DL models trained on these two types of imperfect CSI datasets. Our objective is to assess how our current digital twin aided CSI feedback approach performs in this comparison. In this simulation, we evaluate the spectral efficiency performance and adopt the estimated channel from the UE as the testing data. The estimated channel is obtained through the downlink channel estimation in the target scenario. The BS is configured with an equivalent isotropic radiated power (EIRP) of $43$ dBm, while the UE has a noise figure of $7$ dB. We employ zero forcing \cite{Bjornson2014} to compute the precoding vectors based on the reconstructed channels. Let $\bH_{k} \in \bbC^{N_t \times U}$ denote the stacked user channels at the $k^{\rm{th}}$ subcarrier. The zero forcing precoding vector at the $k^{\rm{th}}$ subcarrier $\bF_{k} \in \bbC^{N_t \times U}$ can be obtained by
        \begin{equation}
            \bF_{k} = \bH_{k}\left(\bH_{k}^{\textrm{H}}\bH_{k}\right)^{-1}.
        \end{equation}
        The sum spectral efficiency can be calculated with \eqref{eq:sum_rate}, where we use the reconstructed estimated channel to design the precoding vectors $\bff_{k, u}$ and the real-world channel $\bh_{k, u}$ for evaluation. \figref{fig:spectral_efficiency} presents the sum spectral efficiency performance with a varying number of users. First, we compare a model trained solely on either real-world estimated CSI or synthetic CSI generated by a digital twin, using 5120 data points for each. The results show that the model trained on the estimated CSI outperforms the one trained on the synthetic CSI. This is expected, as the estimated CSI is derived from real-world measurements and therefore contains more accurate channel information. While the model trained on synthetic CSI achieves a lower spectral efficiency, it requires no real-world data collection. Moreover, the performance gap can be further reduced by refining the model with a small amount of real-world data, as shown in Figure \ref{fig:spectral_efficiency}. Consequently, this highlights the benefits of using digital twins in DL-aided wireless communication tasks, particularly in scenarios where extensive real-world data collection is challenging.

\section{Conclusion} \label{sec:Conclusion}
    In this paper, we have explored the potential of site-specific digital twins in reducing the real-world data collection overhead for DL-aided massive MIMO CSI feedback. We have proposed a novel framework for assessing the fidelity of digital twins, which considers critical aspects including 3D geometry, EM material, ray tracing, and hardware modeling. Moreover, we have developed a refinement approach that utilizes a small amount of real-world data to enhance the model generalization capability, making it adaptable to real-world conditions. Based on the simulation results, we draw the conclusion from the following three aspects:
    \begin{itemize}
        \item \textbf{Site-specific data versus generic data.} Models trained with site-specific digital twins show significantly better performance compared to those trained on generic datasets, highlighting the importance of site-specific data for learning accurate channel representations.
        \item \textbf{Refinement of digital twins.} The proposed online model refinement approach proves effective in enhancing the performance of a model pre-trained on digital twin data using only a small amount of real-world data, showing that digital twins can reduce the need for large-scale data collection while maintaining high accuracy.
        \item \textbf{Fidelity of digital twins.} Our analysis reveals that the fidelity of digital twins, particularly the accuracy of 3D geometry, ray tracing, and hardware modeling, has a significant impact on the quality of CSI reconstruction, with higher fidelity models yielding superior results. The results also show that the accuracy of the materials' EM properties may have less impact compared to other fidelity aspects.
    \end{itemize}
    Overall, this work demonstrates the effectiveness of digital twins in enhancing DL-aided massive MIMO CSI feedback, particularly in scenarios where real-world data collection is challenging or impractical. The proposed approaches pave the way for future research in leveraging digital twins for various wireless communication applications.

\section{Future Directions} \label{sec:Future_Directions}
    Currently, wireless digital twins are still in their infancy, and there is much room for improvement in terms of fidelity and practical applications. In this section, we outline the following directions envisioned for future research in this area:
    \begin{itemize}
        \item \textbf{Twinning fidelity analysis for other wireless communication tasks.} While this work focuses on massive MIMO CSI feedback, future research could investigate how digital twin fidelity affects other tasks, such as beam management~\cite{Jiang2023a}, localization~\cite{Morais2024}, and resource allocation~\cite{Alikhani2024}. Understanding these impacts could facilitate the development of more robust and efficient digital twin models for various wireless communication applications.
        \item \textbf{Alternative refinement data selection methods.} Exploring alternative methods for selecting refinement data is an interesting direction. For instance, LoS and NLoS users may have different sensitivity levels to the digital twin fidelity. LoS channels, which are dominated by a direct path, might be less susceptible to geometric or fidelity variations. Thus, we could consider different data selection strategies for LoS and NLoS users. Additionally, since the main objective of CSI compression and feedback is to help the BS design downlink precoding/beamforming vectors, the UE could leverage the link quality of the downlink data transmission to know the performance of the DL model and decide whether to report the CSI or not, thereby eliminating the need for NMSE computations. Furthermore, more sophisticated methods that leverage additional information, such as geometry or channel distribution statistics, could be explored. However, this may increase the computational overhead on the UE, which is a trade-off that needs to be carefully considered.
        \item \textbf{Extensions to dynamic environments.} In this work, we focused on static environments to solely study the impact of digital twin fidelity on DL-based CSI compression. However, a crucial direction for future research is exploring how digital twins can emulate dynamic environments, where user and scatterer behavior, along with channel conditions, change over time. This will involve developing more sophisticated modeling techniques to account for environmental dynamics and temporal variations. Additionally, ensuring dataset diversity will be a key factor in achieving high performance for DL-aided communication tasks in such dynamic settings.
        \item \textbf{Learnable digital twins.} While this work focuses on physics-based digital twin construction, future research could explore finding a sweet spot between physics-based and data-driven approaches. This hybrid concept, often referred to as learnable digital twins~\cite{Alikhani2025, Jiang2025}, lies in the middle of the spectrum. For example, building upon physics-based digital twins, we could make certain parameters learnable, such as EM material parameters, ray tracing parameters, and hardware model parameters. This adaptive capability would allow the digital twin to dynamically adjust its internal modules based on feedback from the environment. Such an advancement could further improve the efficiency and robustness of digital twin aided wireless communications in diverse and changing environments.
    \end{itemize}

\end{document}